\documentclass[a4paper, 12pt, one column, aas_macros]{article}
\usepackage[english]{babel}
\usepackage[utf8x]{inputenc}
\usepackage[T1]{fontenc}
\usepackage{aas_macros}
\usepackage{color}
\usepackage{xcolor}
\usepackage{graphicx}
\usepackage{epstopdf}
\usepackage{caption}
\usepackage{subcaption}
\usepackage{authblk}
\usepackage{array}
\graphicspath{ {./figure/} }

\newcommand{\oc}[1]{\textcolor{black}{#1}}
\newcommand{\rb}[1]{\raisebox{-0.1ex}{#1}}
\newcommand{\rbd}[1]{\raisebox{-0.3ex}{#1}}
\usepackage[top=1.3cm, bottom=2.0cm, outer=2.5cm, inner=2.5cm, heightrounded,
marginparwidth=1.5cm, marginparsep=0.4cm, margin=2.5cm]{geometry}

\usepackage{graphicx} 
\usepackage[colorlinks=true, linkcolor=blue, urlcolor=blue, citecolor=blue]{hyperref}
\usepackage{amsmath}  
\usepackage{amsfonts} 
\usepackage{amssymb}  

\usepackage[authoryear]{natbib}
\setcitestyle{authoryear,open={(},close={)}}

\title{Bioinspired Drone Rotors for Reduced Aeroacoustic Noise and Improved Efficiency}
\author[1]{Suryansh Prakhar}
\author[1]{Jung-Hee Seo}
\author[1,\footnote{Email address for correspondence: \href{mailto:mittal@jhu.edu}{\texttt{mittal@jhu.edu}}}]{Rajat Mittal}
\affil[1]{Department of Mechanical Engineering, Johns Hopkins University, Baltimore, MD, USA}

\begin{document}
\date{}
\maketitle
\begin{abstract}
The application of unmanned aerial vehicles (UAVs) is surging across several industries, paralleled by growing demand for these UAVs. However, the noise emitted by UAVs remains a significant impediment to their widespread use even though in areas such as product delivery, they can be more environmentally friendly than traditional delivery methods. Nature has often been a source of inspiration for devices that are efficient and eco-friendly.  In the current study, we leverage the previous work by Seo et al. \emph{(Bioinsp. Biomimetics, 16 (4):046019, 2021)} on the aeroacoustics of flapping wing flight in mosquitoes and fruit flies to propose and examine a simple strategy for reducing the aeroacoustic noise from drone rotors. In particular, inspired by these insects, we explore how an increase in the planform area of the rotor could be used to reduce the rotation rate and the associated aeroacoustic noise from small-scale rotors. The study employs a sharp-interface immersed boundary solver for the flow simulations and the aeroacoustic sound is predicted by the Ffowcs Williams–Hawkings equation. Simulations indicate that the simple strategy of employing rotors with larger planform areas could lead not just to reduced aeroacoustic noise but improved power economy as well.
\end{abstract}

\section{Introduction}
Drones are transforming several industries such as transportation, healthcare, vaccine delivery \citep{ziplinecovid}, rescue operations \citep{dronerescue}, food delivery, and environmental monitoring among others. However, the noise they produce during flight is one of the factors preventing their wide-scale use in several applications \citep{DroneNoiseRev}. The authors also list several other downsides such as adverse effects on animals and psychological and psychiatric problems in humans with long-term exposure to such noise.

The dominant source of noise from drones is aeroacoustic noise \citep{DroneNoiseRev} that consists of tonal and broadband \citep{lowmachaerobook} components. Tonal noise is periodic and occurs mainly at the rotation frequency and its harmonics, and generally depends on blade thickness, shape, rotation speed and the advancing/retreating condition of the rotors. Reduction in tonal noise may require changing the shape of the blade or reducing rotational speed but other more unconventional strategies have also been attempted \citep{unevenBladeNoise,UnevenBladeNoiseRed}. Broadband noise can be categorized as noise due to non-periodic fluctuations \oc{and these are generated due to the blade interacting with turbulent flow structures. The broadband noise is particularly important at low tip Mach numbers ~\citep{flight_veh_acoustics} and this can be further sub-categorized as i) noise due to the flow separation, ii) noise due to the interaction of the shed vortices with other blades (blade-vortex interaction, BVI), iii) noise due to the interaction of boundary layer turbulence with trailing edge and leading edge. The incoming flow condition also affects the broadband noise \citep{petricelli2023experimental}. The tonal and the broadband noise together form the total loading noise which is the dominant noise components at low Mach flows \citep{lowmachaerobook}. The aeroacoustics theories~\citep{lighthill1952sound,FWHorigpaper} showed that the source of the loading noise can be expressed by the surface pressure fluctuations.} 
All \oc{the aforementioned} flow structures affect the surface pressure of the blade and generate loading noise. Computational studies on drone rotors show complex vortical structures \citep{droneflowstruct} and the interaction of the blade with these vortical structures is one of the major sources of the broadband noise as mentioned above. There have been several design modifications such as modifying the blade tip, the leading or trailing edge, blade planform shape and so on to try to mitigate the blade-vortex interaction noise \citep{yu_bvi}. In some studies, bio-inspired designs have been employed to reduce the noise. The flight of the owls has been studied so that their salient features can be used to design silent UAVs. For instance, the wings of an owl contain features such as serrations and fringes to promote silent flight \citep{wagner_owl_wing} which was used by \cite{Hao_owl} to develop a quieter drone propeller. 

\begin{figure}
\centering
\includegraphics[width=\textwidth]{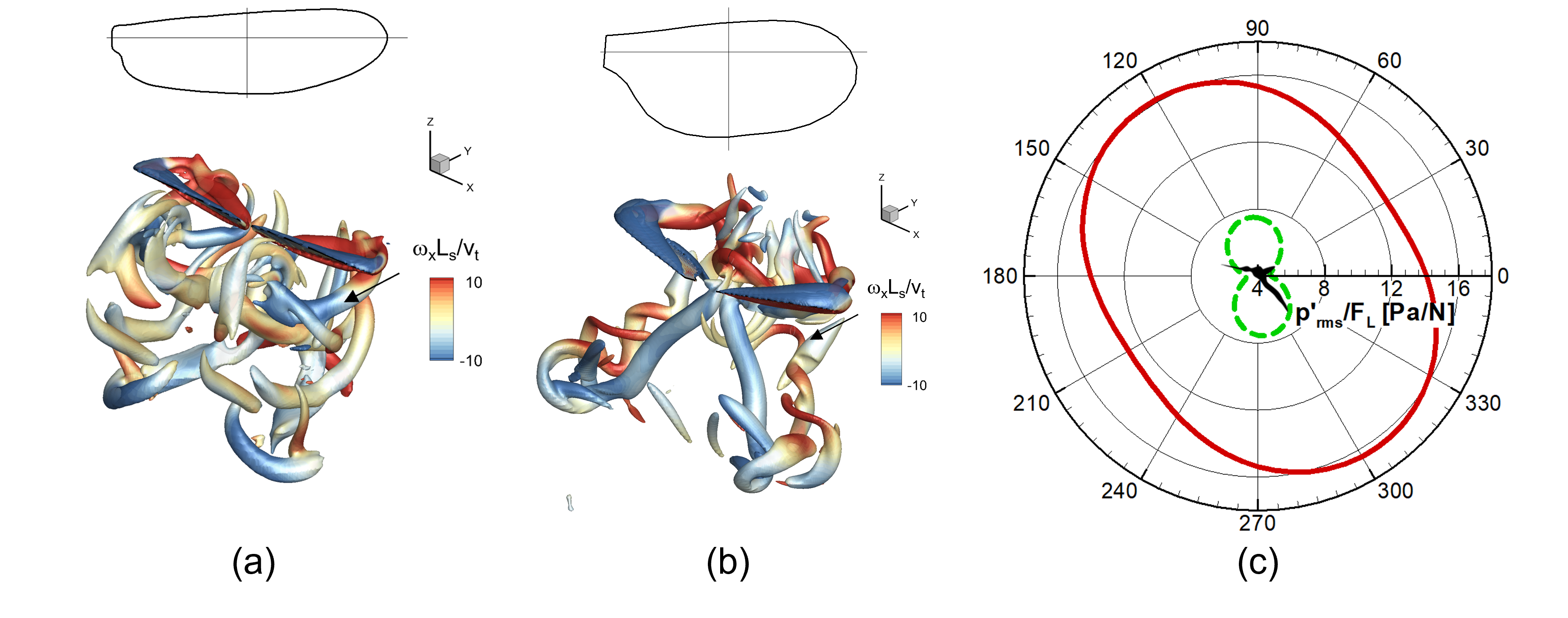}
\caption{The results from \cite{Seo_mosq} showing the wing shapes and vortex structures for  the (a) mosquito and (b) the fruit fly. (c) The directivity pattern comparing the noise per lift generated by both these insects.}
\label{fig:comparition:mosquito_and_ffly}
\end{figure}
The work in this paper is inspired by a study on the aeroacoustics of flapping flight in insects \citep{seo_mos_scale,Seo_mosq} where the authors compared the sound generated by  flapping of mosquito and fruit fly wings. Despite comparable wing spans and weights,  mosquitoes are known to generate a higher intensity sound from their flapping than fruit flies, and these studies examined the features of the wing morphology and kinematics that might explain this difference. The study employed direct numerical simulations of the flow coupled with an estimation of far-field sound via the Ffowcs William-Hawkings (FWH) acoustic analogy-based model and figure \ref{fig:comparition:mosquito_and_ffly}(a-b) shows the vortex structures for the two wings. Also shown in figure \ref{fig:comparition:mosquito_and_ffly}c is a comparison of the directivity pattern of the radiated aeroacoustic sound in terms of root-mean-square acoustic pressure normalized by the lift generated by the wing and the mosquito clearly generates a higher acoustic pressure per unit lift  than the fruit fly. Indeed, as shown in the table \ref{table:comparition}, the specific acoustic power of the mosquito is about 3.7 times larger than a fruit fly.

The study then compared the relevant attributes of the wing and the flapping kinematics in order to determine the features that are responsible for this increased specific acoustic power. The primary factor was found to be the significantly higher flapping frequency of the mosquito (720 Hz) compared to the fruit fly (218 Hz) and this was inline with the fact that far-field acoustic pressure scales with the cube of the flapping frequency. This brings up the question: how does the fruit fly then generate a similar magnitude of lift as the mosquito with a much lower flapping frequency from wings that have a similar total wing-span? The study showed that while the wing spans of the two insects are similar, the wings of a fruit fly have a much larger area (almost 57\% larger) than the fruit fly. There are also other differences as well including a larger stroke amplitude for the fruit fly and a lower aspect ratio wing planform, but it is primarily the larger ``wetted'' area of the fruit fly wing that allows the fruit fly to generate a lift equivalent to the mosquito despite having a much lower flapping frequency. Interestingly, it was also shown that despite the significant differences in sound power, the overall specific aerodynamic power (power required for flapping per unit lift) was also very similar (see table \ref{table:comparition}), suggesting that these differences between the fruit fly and the mosquito are achieved without significantly affecting the power efficiency for flight.
\begin{table}
\centering
\begin{tabular}{|l|c|c|}
\hline
\textbf{Attribute} & \textbf{Mosquito} & \textbf{Fruit Fly} \\
\hline
Body weight (mg) & 0.81 & 1.22 \\
\hline
Wing area (mm$^2$) & 1.85 & 2.9 \\
\hline
Wing span (mm) & 2.75 & 2.86 \\
\hline
Flapping frequency (Hz) & 720 & 218 \\
\hline
Wing aspect ratio & 4.2 & 2.15 \\
\hline
Maximum wing tip velocity (m/s)& 5.13 & 5.33 \\
\hline
Mean lift ($\mu$N)& 8 & 12 \\
\hline
Mechanical power per unit lift (W/N)& 2.62 & 2.5 \\
\hline
Acoustic power per unit lift($\mu$W/N)& 3.32 & 0.91 \\
\hline
\end{tabular}
\caption{Comparison of mosquito and fruit fly parameters and some key results with the input kinematics and the wing planform data based on \cite{beatus_wingK,bomphrey_wingK,zhang_wingK} and the results are based on \cite{Seo_mosq}.}
\label{table:comparition}
\end{table}

\begin{figure}
\centering
\includegraphics[width=\textwidth]{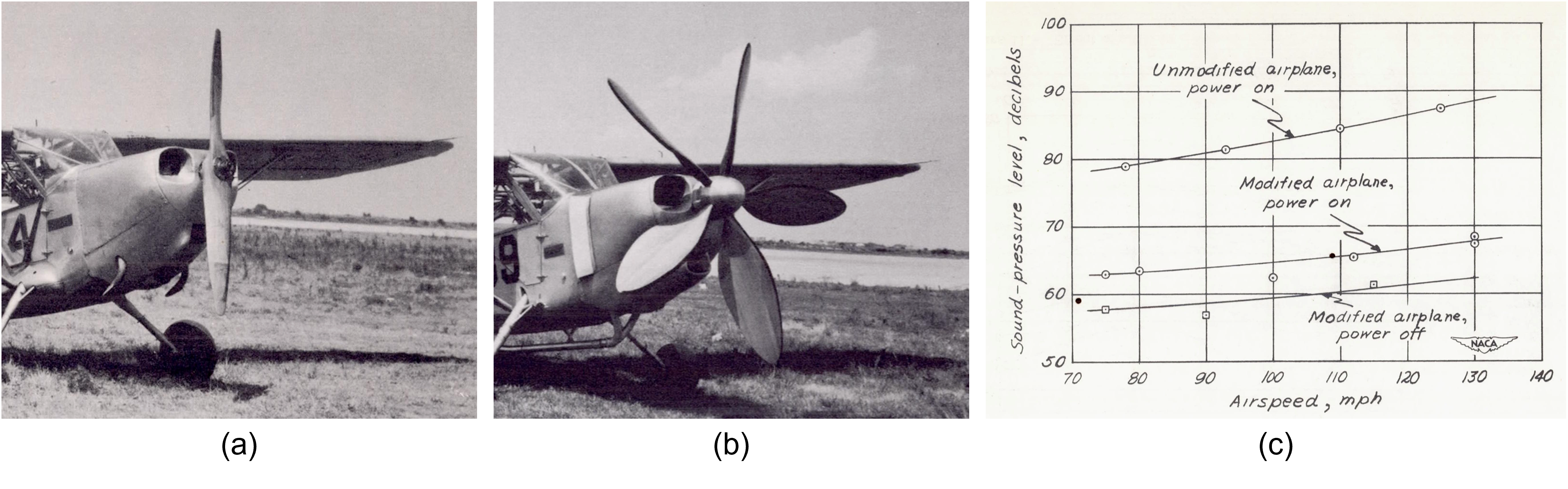}   
\caption{(a) and (b): Comparison of the airplane nose propeller used by \cite{vogeley1949} and we note that the modified nose propeller have number of blades increased to five with the area of each blade and the propeller diameter increased as well. (c) Comparison of the sound pressure level at various airspeeds. All images are taken from their report.}
\label{fig:vogeley_exp}
\end{figure}

Thus, this comparative analysis suggests that an increase in the wetted area of the rotor blades accompanied by a reduction in the rotation speed could be an effective strategy for reducing the aeroacoustic sound from rotors for small-scale drones. Indeed, there is evidence that such a strategy could be useful even for larger-scale rotors. More than 70 years ago \cite{vogeley1949} demonstrated a reduction in aeroacoustic sound from an aircraft propeller using a similar idea. Figure \ref{fig:vogeley_exp} taken from this study  shows the original two-blade nose propeller (figure \ref{fig:vogeley_exp}a), which is replaced by a five-bladed propeller (figure \ref{fig:vogeley_exp}b) where each propeller blade has a significantly larger wetted area and the radius of this propeller was also increased. Thus at a given airspeed of 55 mph, the modified blade propeller operated at a reduced rotation rate (794 RPM) compared to the unmodified propeller (2130 RPM) and this resulted in a reduction in the overall level of the aeroacoustic sound. Figure \ref{fig:vogeley_exp}c shows a minimum reduction of about 15 dB in noise associated with this modification of the propellers and the authors mention that  at highest airspeed and a given sound threshold of 65 dB, the unmodified plane has to be about 2000 ft away from the observer, while the modified plane could come as close as 200 ft. Interestingly, the modified aircraft also exhibited a  higher mechanical efficiency, for instance, the mechanical efficiency with the modified propeller was 7.2\% higher at an airspeed of 55 mph. These results suggest that similar design changes in drone rotors could lead to a reduction in acoustic emissions for a given lift while maintaining or even improving the power efficiency. This could result in quieter drones that are environmentally friendly and even increase the flight duration through improved efficiency.

In the current study, we employ direct numerical simulations (DNSs) of the flow coupled with computation of far-field sound to examine the proposition that an increase in the wetted area of a small-scale rotor blade could be used to reduce the far-field sound for a given amount of lift force.

\section{Methodology}
\subsection{Rotor Configuration}
We examine rotors in hover, and the majority of the study is for a single rotor blade with a limited number of simulations for a two-bladed rotor. The baseline configuration is based on the experiments of \cite{45RotorExp}, who employed a thin, flat rectangular blade with an aspect-ratio of 5 with the inner edge at a distance of $R_i/R_o =0.244$ from the axis of rotation, where $R_o$ is the outer radius of the rotor blade. The second configuration (BI-13) is also a thin flat rotor with a wetted area that is 30\% larger than the baseline rotor. It has the same span, but a shape that is inspired by both the elliptic-shaped wing of fruit flies and the modified propeller blade used in the study of \cite{vogeley1949}. The third and fourth configurations (BI-20 and BI-40) are based on BI-13, except they have areas that are 2$\times$ and 4$\times$ the baseline rotor, respectively. 
\begin{figure}
    \centering
 \begin{subfigure}[b]{0.24\textwidth}
      \centering
         \includegraphics[width=\textwidth]{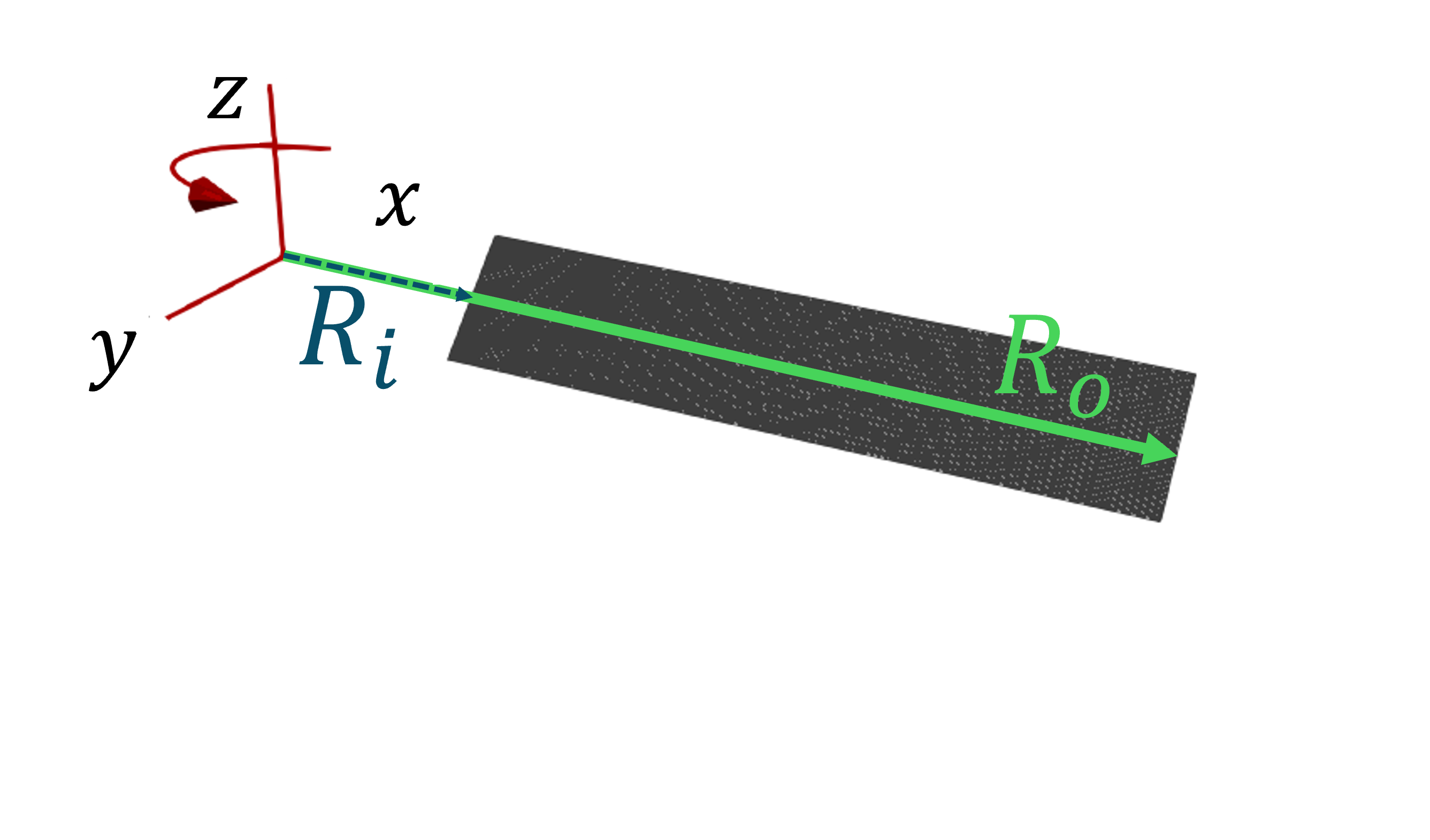}
         \caption{}
         \label{fig:geo:wing:a}
     \end{subfigure}
  \hfill
     \begin{subfigure}[b]{0.24\textwidth}
         \centering
         \includegraphics[width=\textwidth]{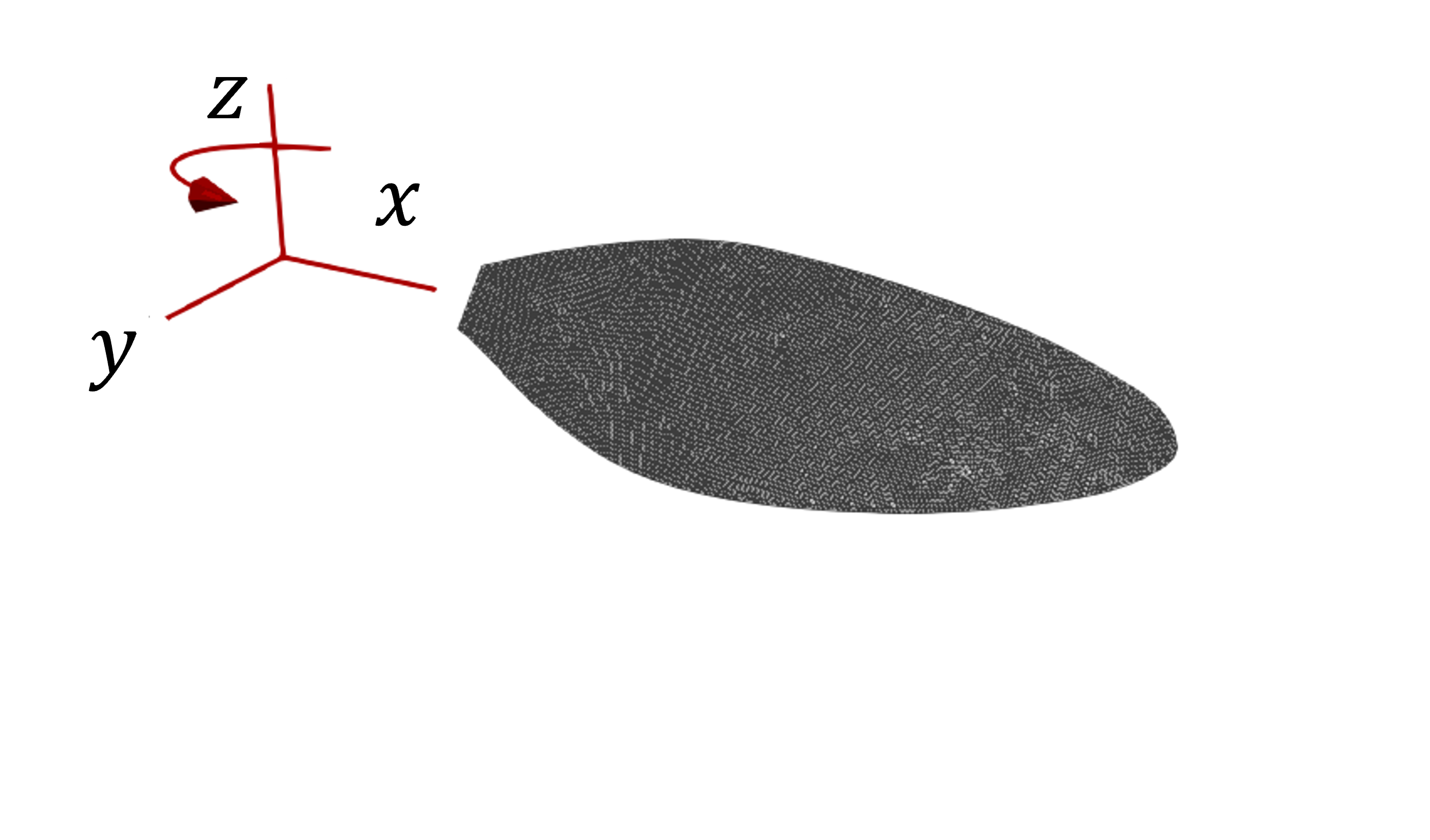}
         \caption{}
         \label{fig:geo:wing:b}
     \end{subfigure}
   \hfill
     \begin{subfigure}[b]{0.24\textwidth}
         \centering
         \includegraphics[width=\textwidth]{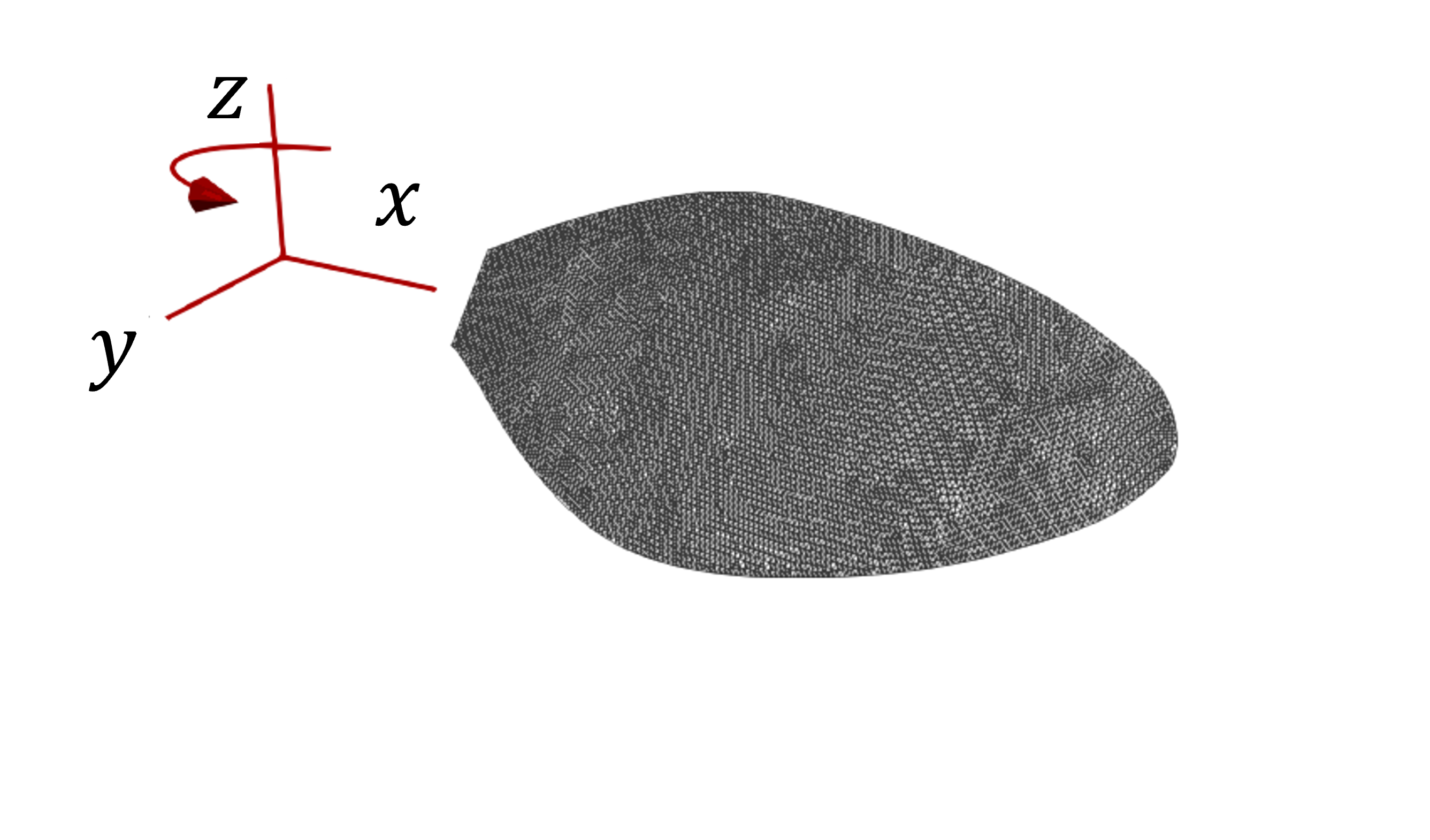}
         \caption{}
         \label{fig:geo:wing:c}
     \end{subfigure}
     \hfill
     \begin{subfigure}[b]{0.24\textwidth}
         \centering
         \includegraphics[width=\textwidth]{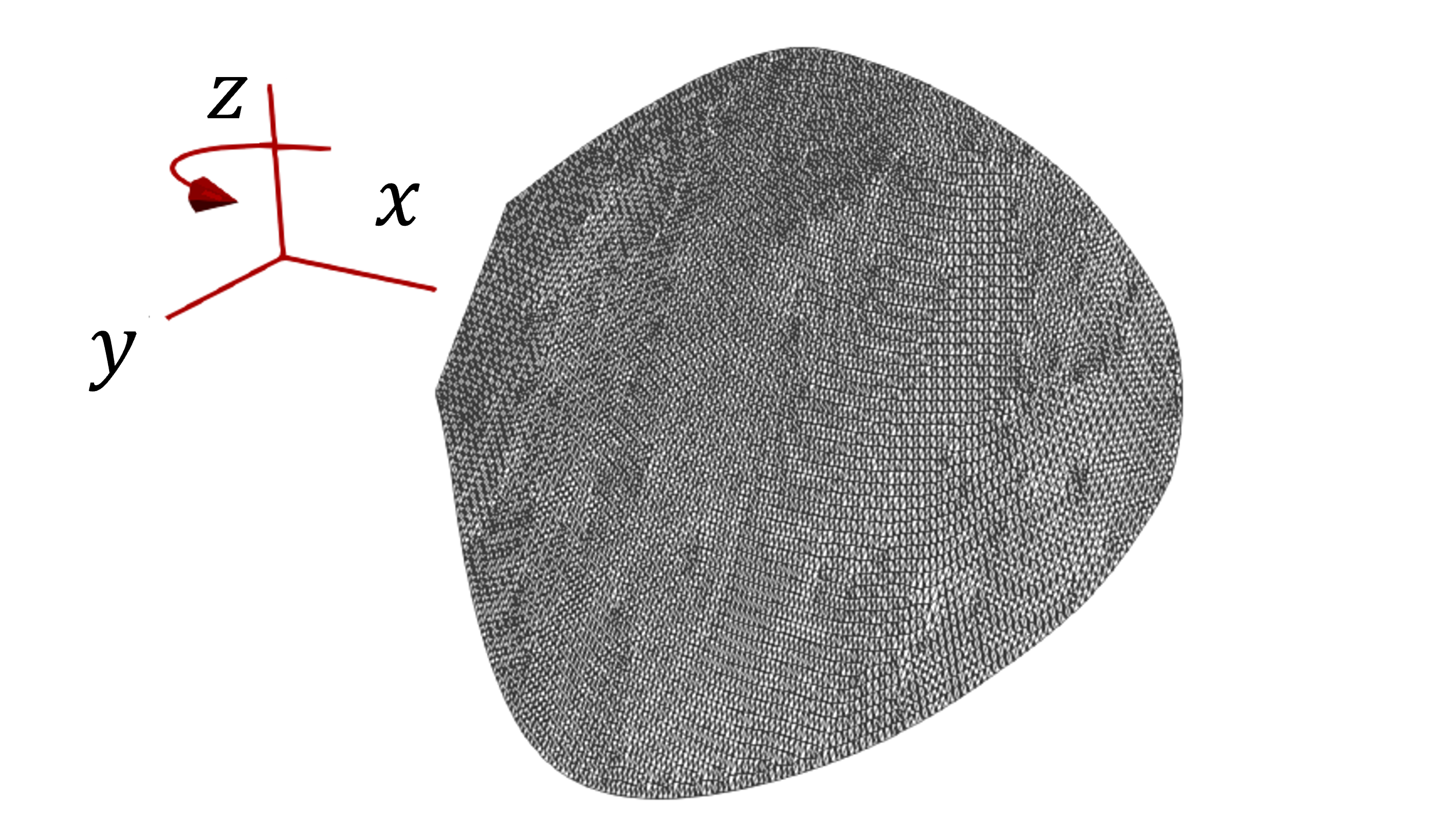}
         \caption{}
         \label{fig:geo:wing:d}
     \end{subfigure}
        \caption{Rectangular and bio-inspired rotor showing (a) Rectangular blade RR-10, (b) Bio-inspired blade with 30\% increased area - BI-13, (c) Bio-inspired blade with 100\% increased area - BI-20 and (d) Bio-inspired blade with 300\% increased area - BI-40. The axis shown at center of rotation and the ratio of the number at the end of the rotors name represents the area ratio between rotors.}
\label{fig:geo:wing}
\end{figure}

In addition to the shape of the blade, the aerodynamics and the performance of these rotor blades are a function of two non-dimensional parameters: the blade pitch-angle $\alpha$ and the non-dimensional rotation speed defined as $\tilde{\Omega}=\Omega/\left( \nu/R_o^2 \right)$, where $R_o$ is the tip radius of the rotor blade. This non-dimensional number is also identical to the tip-velocity based Reynolds number and we note that for a small-scale drone rotor of radius $R_o$=10 cm rotating at 3000 RPM, $\tilde{\Omega} \approx 200,000$.

\subsection{Flow Solver}
\begin{figure}
\centering
\includegraphics[width=0.7\textwidth]{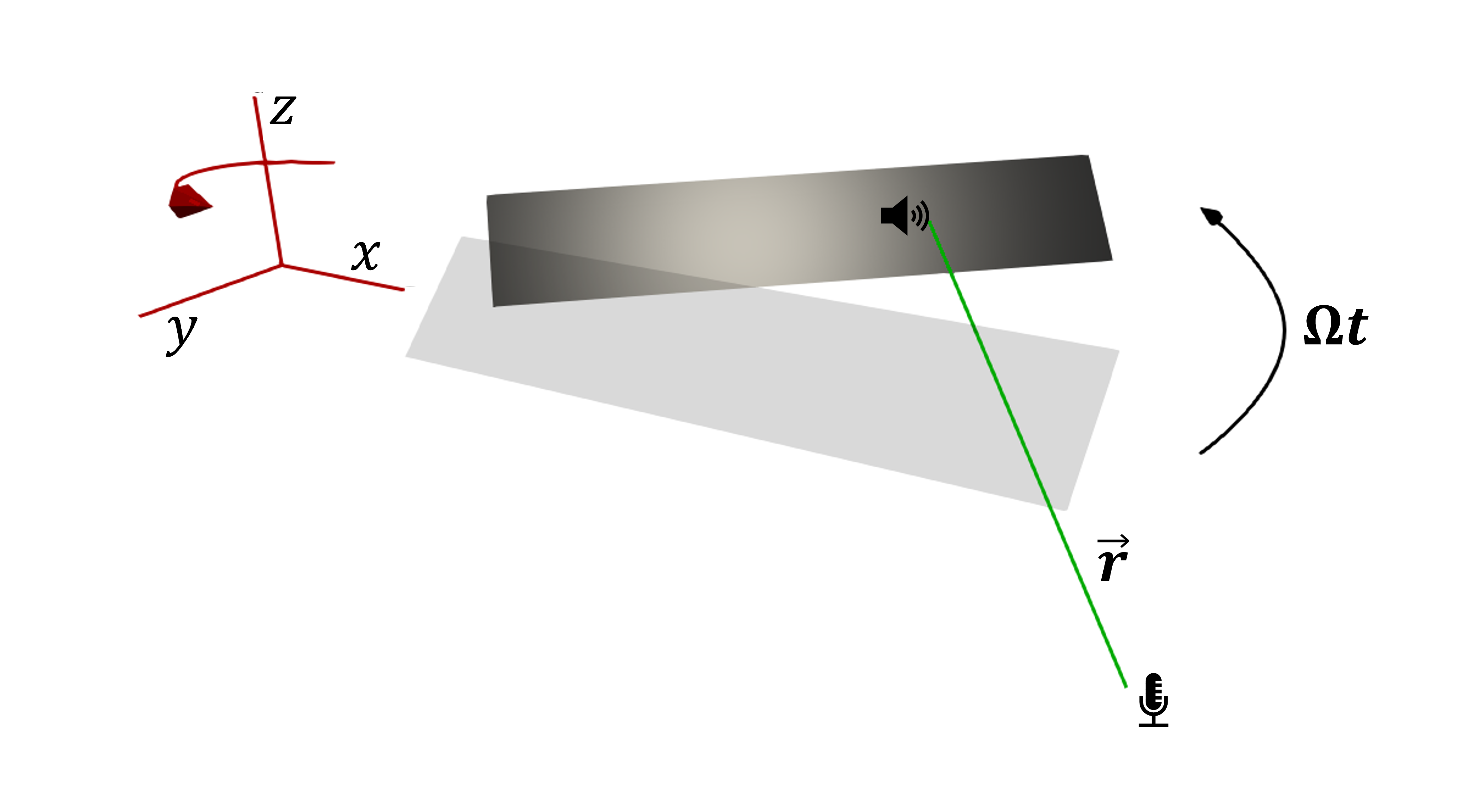}
\caption{A snapshot of the rotor at time $t$ in non-inertial rotating reference frame (semi-transparent) for flow solver and after being shifted to the ground frame (black) for FWH based acoustic solver. The axis is shown at the center of rotation with the lift force along z-axis and the span of rotor along the x-axis (when rotor is in reference frame). The microphone icon represents a sample recording location (not to scale).}
\label{fig:fwh_rrf}
\end{figure}
 We use our in-house Navier-Stokes solver called ViCar3D \citep{mittal2008,seo2011} to solve the 3D incompressible Navier-Stokes equations in non-inertial rotating reference frame \citep{batchelor_book,speziale_rrf_eqn} given by 
 \begin{equation}
     \nabla \cdot {\bf u} = 0 \,\,\, ,
     \label{conteqn}
 \end{equation}
 and
 \begin{equation}
     \frac{\partial {\bf u}}{\partial t} + {\bf u}\cdot \nabla {\bf u}  = -\frac{1}{\rho}\nabla P +\nu \nabla^2 {\bf u} -2{\bf \Omega} \times {\bf u} - {\bf\Omega \times (\Omega \times x)} \,\,\,,
\label{momeqn}
 \end{equation}
 where, ${\bf\Omega} = \left(  0, 0, \Omega \right) $ is the rotation vector of the reference frame, $2{\bf \Omega} \times {\bf u}$ is the Coriolis acceleration term and ${\bf\Omega \times (\Omega \times x)}$ is the centrifugal acceleration term. The tip Mach numbers for small-scale drone rotors (say 10 cm radius at a 3000 RPM) are in the range of 0.09, and therefore, the incompressible assumption is reasonable for this study.
 
The rotor blade is represented by a zero-thickness plate comprised of unstructured triangular elements while the flow domain is discretized  using a non-uniform Cartesian grid.  The momentum equation uses Van-Kan's fractional step method to decompose it into an advection-diffusion equation and a pressure Poisson equation. A second-order Adam-Bashforth scheme is used for the advection-diffusion equation, an implicit Crank-Nicholeson scheme is used for the viscous term, and biconjugate stabilized gradient descent is used for the pressure Poisson equation.

\subsection{Acoustics}
To compute the noise generated by the rotors, we use the Ffowcs Williams-Hawkings (FWH, \cite{FWHorigpaper}) equation and to make the computation less expensive, we specifically use the Brentner and Farassat integral formulation of the FW-H equation \citep{fwheqn2,brentner_fwh_term_def} given by 
\begin{equation}
\begin{split}
        4\pi p^{\prime}({\bf x},t) = & \frac{1}{c}\int \Big[\frac{\dot{L}_r}{r(1-M_r)^2} \Big]_{t-r/c} dS
    +\int \Big[\frac{L_r-L_M}{r^2(1-M_r)^2} \Big]_{t-r/c} dS \\
    & +\frac{1}{c}\int \Big[\frac{L_r(r\dot{M_r}+c(M_r-M^2))}{r^2(1-M_r)^3} \Big]_{t-r/c} dS \,\, ,
    \label{fwheqn}
    \end{split}
\end{equation}
where $p^\prime$ is the sound pressure, $c$ is the speed of sound, $r$ is the distance between surface to monitoring point, $L_r={\bf L}\cdot {\bf r}$ with ${\bf r}$ being a  unit normal pointing from the surface to the monitoring point and $L_M={\bf L}\cdot {\bf M}$. Here, ${\bf L}$ is the force vector given by $P{\bf n}$ with ${\bf n}$ being the surface normal vector.  The Mach number vector, ${\bf M}$ is calculated as ${\bf v}/c$ where v is the surface velocity and the Mach number of the surface in the direction of the monitoring point ($M_r$) is calculated using ${\bf M}\cdot {\bf r}$. 

To capture both the tonal and the broadband noise, the body surface is first multiplied by a transformation matrix to shift it back to the ground frame (shown in figure \ref{fig:fwh_rrf}) along with the rotor surface velocities adjusted to be with respect to the ground frame and these values were then used to calculate the sound pressure at a given location. 
We also calculate the directivity by calculating the RMS value of the pressure signal ($p^{'}_{rms}$)
and the acoustic intensity using 
\begin{equation}
    I_a=  \frac{1}{4 \pi r^2 \rho c} \int_S (p_{rms}^{'})^2 dS \,\,\, .
    \label{AcousticInt}
\end{equation}
\color{black}
\subsection{Solver Validation}
\begin{figure}
\centering
\includegraphics[width=0.9\textwidth]{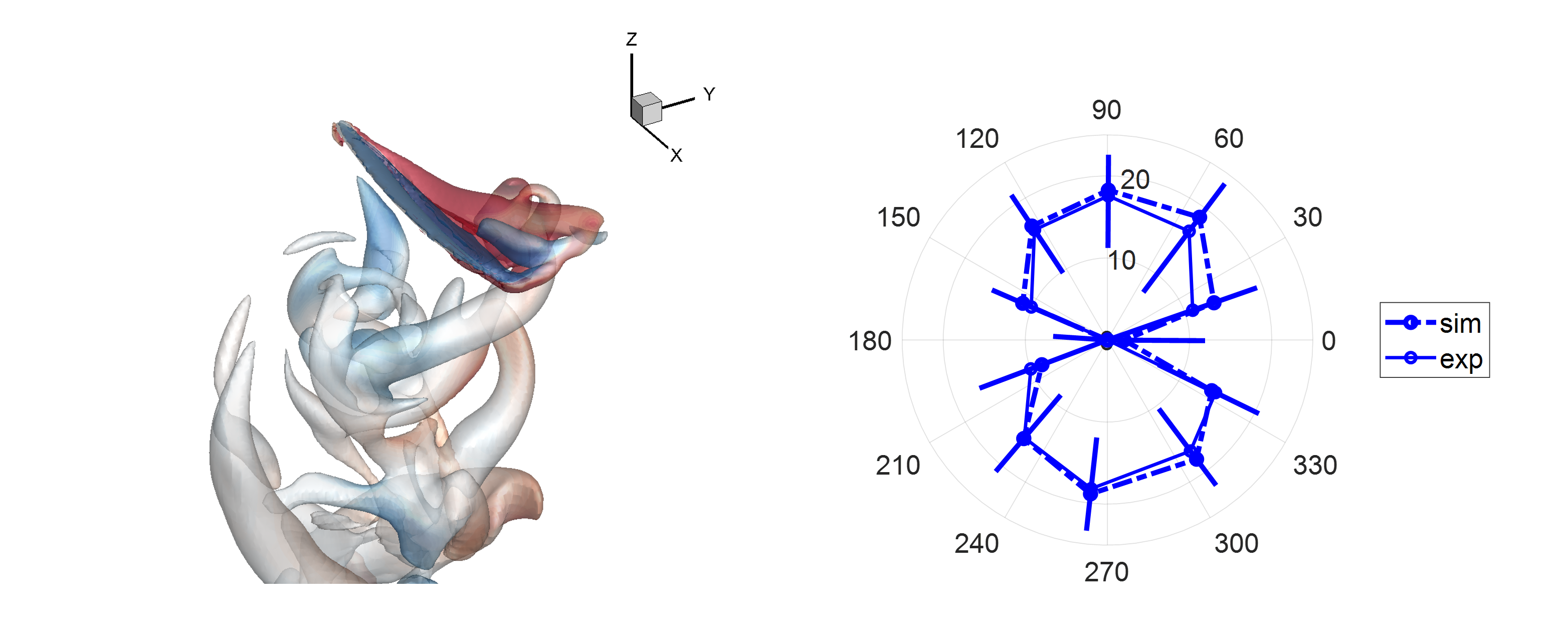}
\caption{\oc{A validation study of ViCar3D flow and  acoustic solver showing the vortices produced by the mosquito wing, and comparing the experimental and computational results of mosquito wing sound directivity pattern. The results are taken from \cite{mosquito_exp_val}.}}
\label{fig:mosquito_exp_validation}
\end{figure}

\cite{seo2022} simulated a canonical case of circular cylinder at Re=100 with Mach number of 0.1 with this same solver and compared the results of FWH based acoustic solver for far field noise with direct acoustic numerical simulation (DaNS) showing an excellent match between the two.
The solver has also been validated for the wing tones generated by the mosquito wing. \cite{mosquito_exp_val} recorded the wing tones from a tethered mosquito. The wing motion was reconstructed from high-speed videos and the flow was simulated by ViCar3D and acoustics was predicted by using the FWH equation. The sound directivity pattern from the simulation and the experiment is reproduced from \cite{mosquito_exp_val} in Fig. \ref{fig:mosquito_exp_validation}. The comparison is quite reasonable in terms of overall intensity and directivity thereby providing confidence in the simulation approach. 
The reader can find additional details and validation case studies for the ViCar3D solver in \cite{mittal2008,seo2011} and specific applications involving rotating and flapping wing can be found in \cite{FlapVsRotWing,new_ibm}. The same flow-acoustic solver is used for the current rotor simulations. While we do not have experimental data for comparison and validation at the relatively low Reynolds that we are limited to in the current study, the above validation for similar flows combined with a grid verification study provides confidence in the accuracy of the results presented here.

\color{black}
\subsection{Scaling Analysis}
\label{scaling}
The underlying assumption in this relatively simple scaling analysis is that the aerodynamic forces are primarily due to pressure, and shear induced forces can be neglected at these relatively high Reynolds numbers. The pressure lift and drag are computed as follows:
\begin{equation}
\begin{split}
    F_L= -\int_S P n_z dS = -n_z \int_S P dS =F_N \sin{\theta} \\
    F_D= -\int_S P n_y dS = -n_y \int_S P dS =F_N \cos{\theta} =F_L \tan{\theta}
\end{split}
    \label{lifteqn}
\end{equation}
where $S$ is the surface of the body and $n_z$ is the outward pointing normal from the blade in the direction of lift. Furthermore, $\theta$ is the pitch angle of the rotor blade. The lift force ($F_L$) for a rotor can be expressed as,
\begin{equation}
    F_L= \frac{1}{2} \rho v_{t}^2 A  C_L  = \frac{1}{2} \rho R_o^2 \Omega^2 A  C_L \,\, ,
    \label{eqn:forcescale}
\end{equation}
where $C_L$ is the lift coefficient, $v_t$ is the tip velocity and $A$ is the surface area of the rotor.

The mechanical power required for the blade to work against the pressure loading (referred to as aerodynamic power hereinafter) is given by
\begin{equation}
    W_M = \int_S P \left( {\bf n} \cdot {\bf v} \right) dS \,\,\, ,
    \label{mechpower}
\end{equation}
where ${\bf v}$ is the surface velocity vector and ${\bf n}$ is the surface normal. In the current study, $\bf{v}= \bf{r} \times \Omega \hat{k}$, where $\bf{r}$ is the position vector of the surface element. Since the blade velocity is in the direction of the drag force on the blade, this aerodynamic power scales as:
\begin{equation}
    W_M \sim F_D R_o \Omega  \sim F_L  R_o \Omega \tan{\theta} \sim R_o^3 \Omega^3 A  C_L    \tan{\theta}.
\end{equation}
\oc{where $F_D R_o$ corresponds to the aerodynamic torque on the blade.}

Similarly, for acoustic intensity (and also for acoustic power), we first need to calculate the scaling of the sound pressure ($p'$). The tonal noise for a rotor in hover is associated primarily with the rotation of the drag vector with the rotor blade and the scaling for this can be inferred from the expression in Eq. \ref{fwheqn} \citep{seo_mos_scale} as,
\begin{equation}
    p' \sim \frac{dF}{dt} \sim F_D \Omega \sim  R_o^2 \Omega^3 A  C_L    \tan{\theta}  \,\,\,.
\end{equation}
Using equation \ref{AcousticInt}, the acoustic power and intensity will scale as square of the sound pressure and thus,
\begin{equation}
    I_a \sim p'^2 \sim R_o^4 \Omega^6 A^2  C^2_L    \tan^2{\theta}  \,\,.
    \label{eqn:int_sp_scaling}
\end{equation}

The scaling in Eq. \ref{eqn:forcescale} indicates that for a constant span $R_o$ and a fixed lift force $F_L$ generated by the rotor \oc{(which also implies a fixed disk loading)}, the rotor rotation speed has to satisfy the following relationship:
\begin{equation}
 \Omega^2 \sim \frac{1}{A}  \frac{1}{C_L}\,\, .
\end{equation} 
Combining the above with the expression in Eq. \ref{eqn:int_sp_scaling}, we find that the acoustic intensity for a fixed lift \oc{(and disk loading)} would scale as:
\begin{equation}
    I_a \sim \frac{1}{A}  \frac{1}{C_L}\,\,\, .
    \label{eqn:intensity_scaling}
\end{equation}
This indicates that for the condition of fixed value of lift \oc{(and also a fixed disk loading since the rotor disk area is the same for all the rotors here)}, the acoustic intensity reduces inversely with the blade area. Substituting the above scaling in the expression for aerodynamic power, we obtain the following:
\begin{equation}
    W_M \sim A^{-\frac{1}{2}} C_L^{-\frac{1}{2}}  \tan{\theta}.\,\,\,\,
    \label{eqn:power_scaling}
\end{equation}
The above suggests that the aerodynamic power would also reduce with an increase in blade area for a constant lift, albeit at a slower rate of $1/A^{\frac{1}{2}}$. 

The reduction in aeroacoustic noise and aerodynamic power with increased blade area however rests on the assumption that $C_L$ is not adversely affected by the increase in blade area. We note here that in general, the lift coefficient of the rotors under consideration depends on the rotation speed, blade shape, and blade pitch. In principle, several non-dimensional parameters might be needed to characterize the shape of the blade but given the relatively simple shapes of the blade, the aspect-ratio $\zeta = (R_o-R_i)^2/A$ is expected to be the dominant shape parameter. Thus, the dependence of $C_L$ can be expressed as: 
\begin{equation}
    C_L=C_L \left( \tilde{\Omega}, \zeta ; \theta \right). \,\,\,\,\, 
    \label{eqn:CL_scaling}
\end{equation}
Thus, it remains to be seen how well the $C_L$ is preserved by changes in blade area and we will examine this in the next sections.

\section{Results}
In these first two sections, we examine the effect of increasing blade area as a strategy for reducing the aeroacoustic noise and aerodynamic power of the rotor for fixed $F_L$. 

\subsection{Blade Pitch \texorpdfstring{$\theta=45^\text{o}$}{theta45}}
\label{results45AoA}
\begin{figure}
    \centering
 \begin{subfigure}[b]{0.24\textwidth}
      \centering         
      \includegraphics[width=\textwidth]{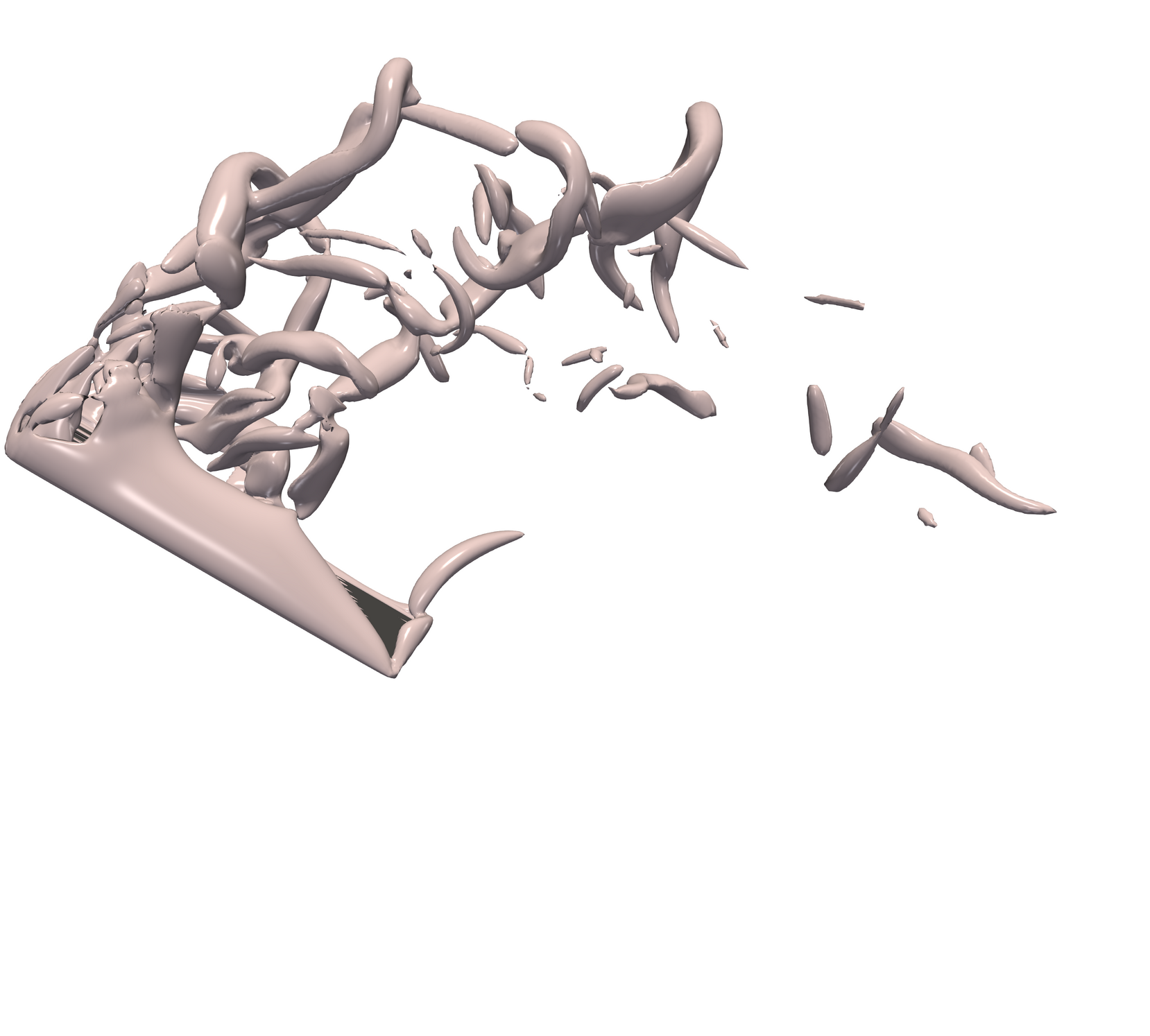}
         \caption{}
         \label{fig:45_Qcrit_all_case:a}
     \end{subfigure}
  \hfill
     \begin{subfigure}[b]{0.24\textwidth}
         \centering
         \includegraphics[width=\textwidth]{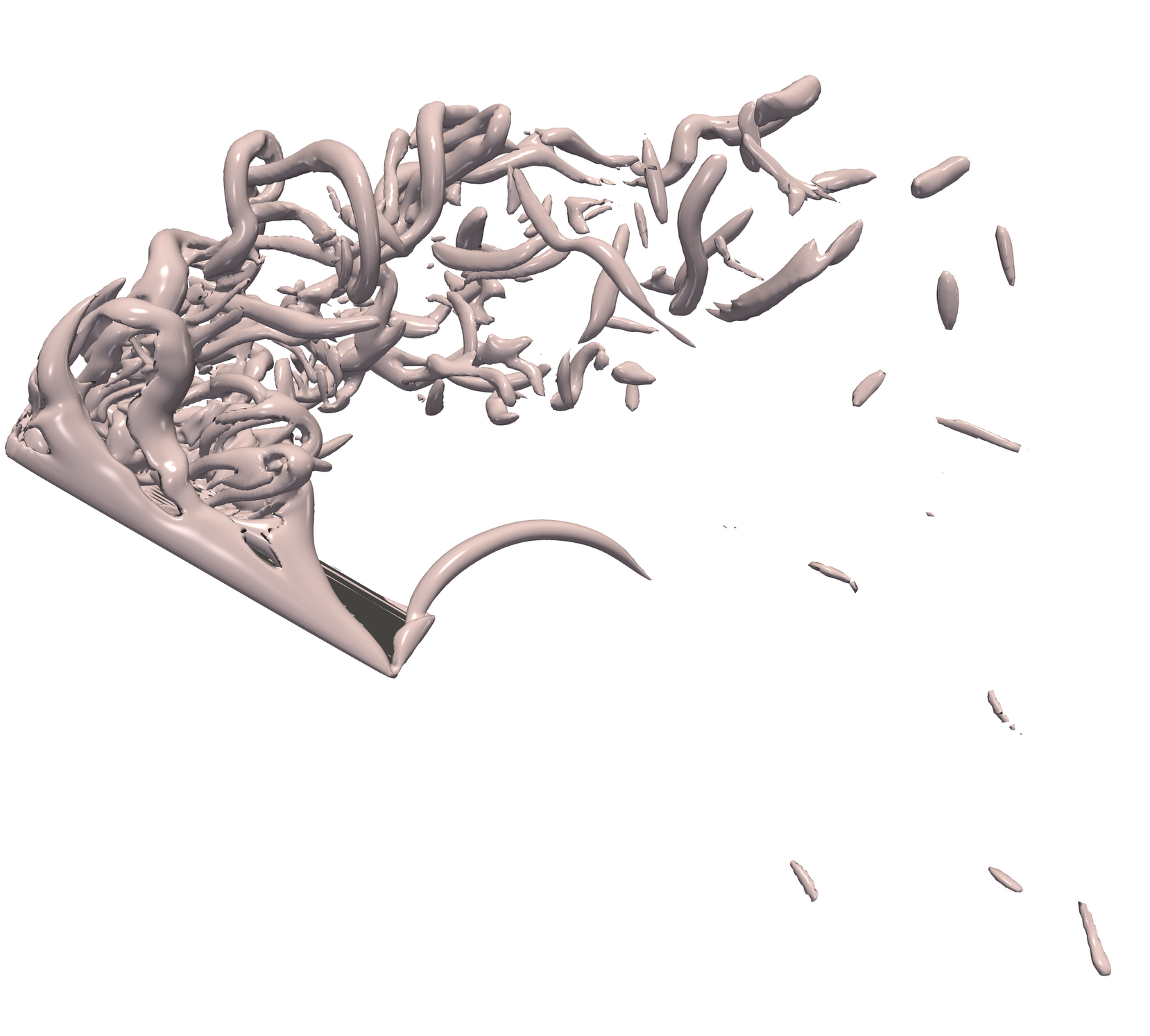}
         \caption{}
         \label{fig:45_Qcrit_all_case:b}
     \end{subfigure}
   \hfill
     \begin{subfigure}[b]{0.24\textwidth}
         \centering
         \includegraphics[width=\textwidth]{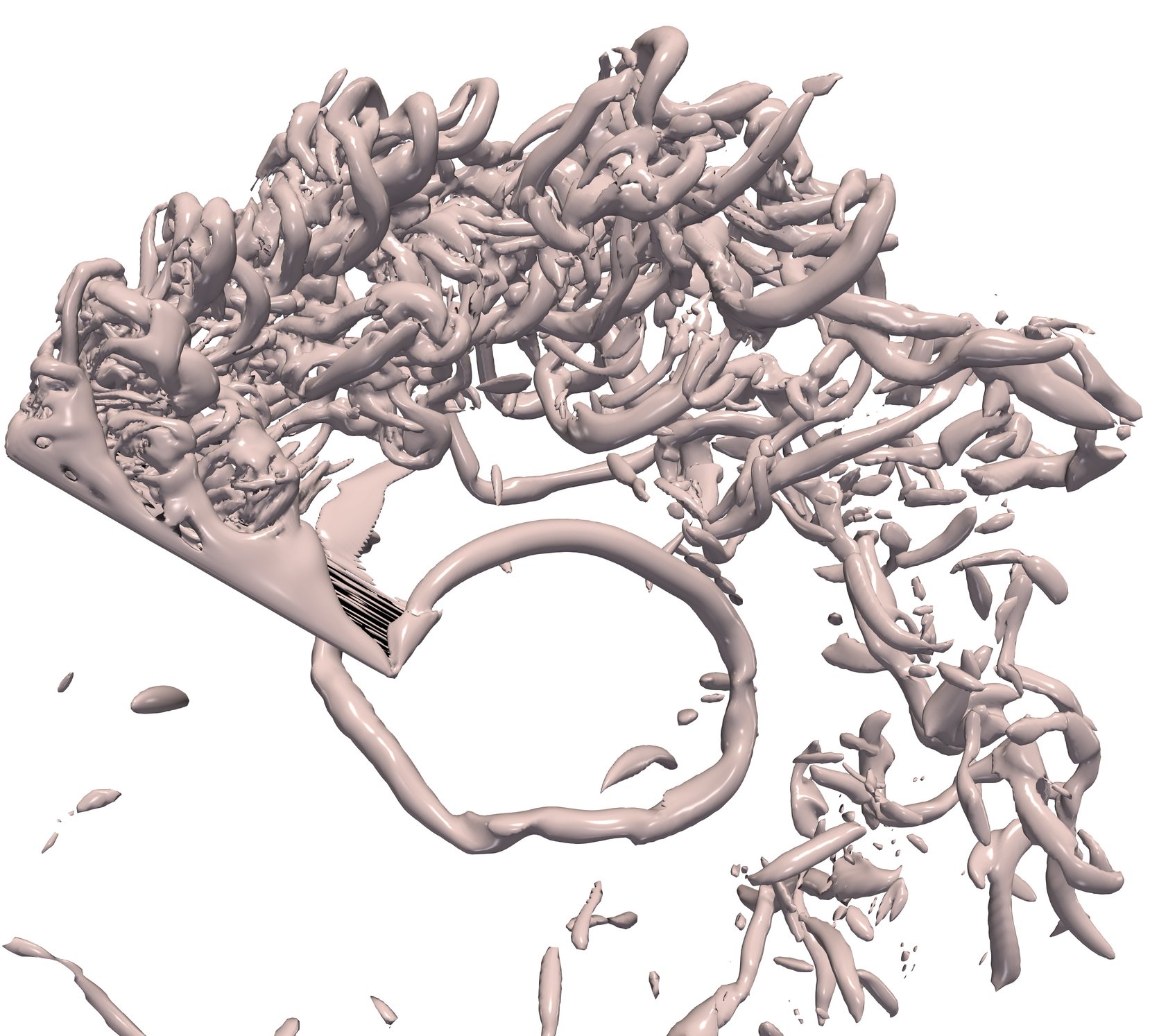}
         \caption{}
         \label{fig:45_Qcrit_all_case:c}
     \end{subfigure}
     \hfill
     \begin{subfigure}[b]{0.24\textwidth}
         \centering
         \includegraphics[width=\textwidth]{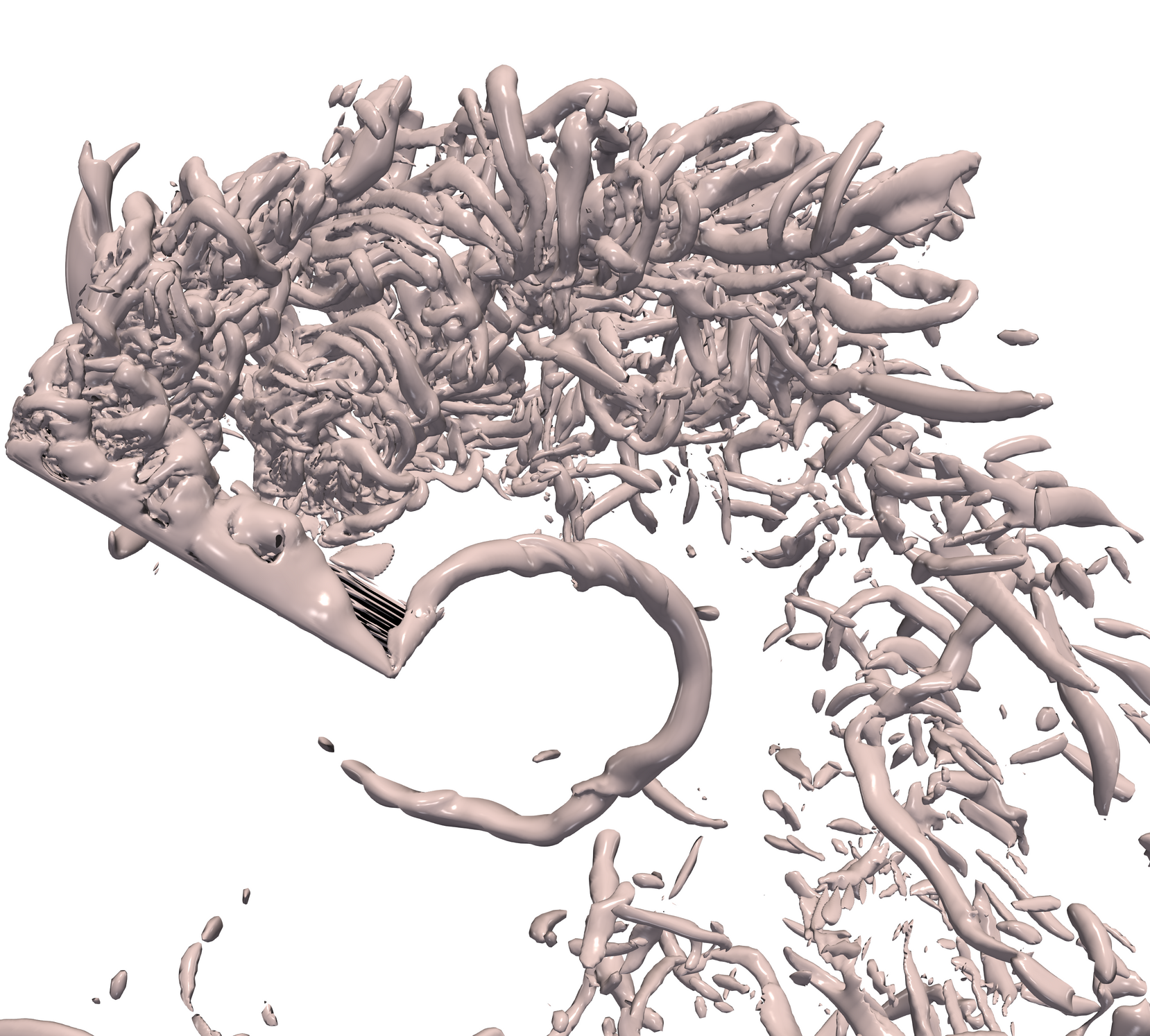}
         \caption{}
         \label{fig:45_Qcrit_all_case:d}
     \end{subfigure}
     \hfill
     \begin{subfigure}[b]{0.6\textwidth}
         \centering
         \includegraphics[width=\textwidth]{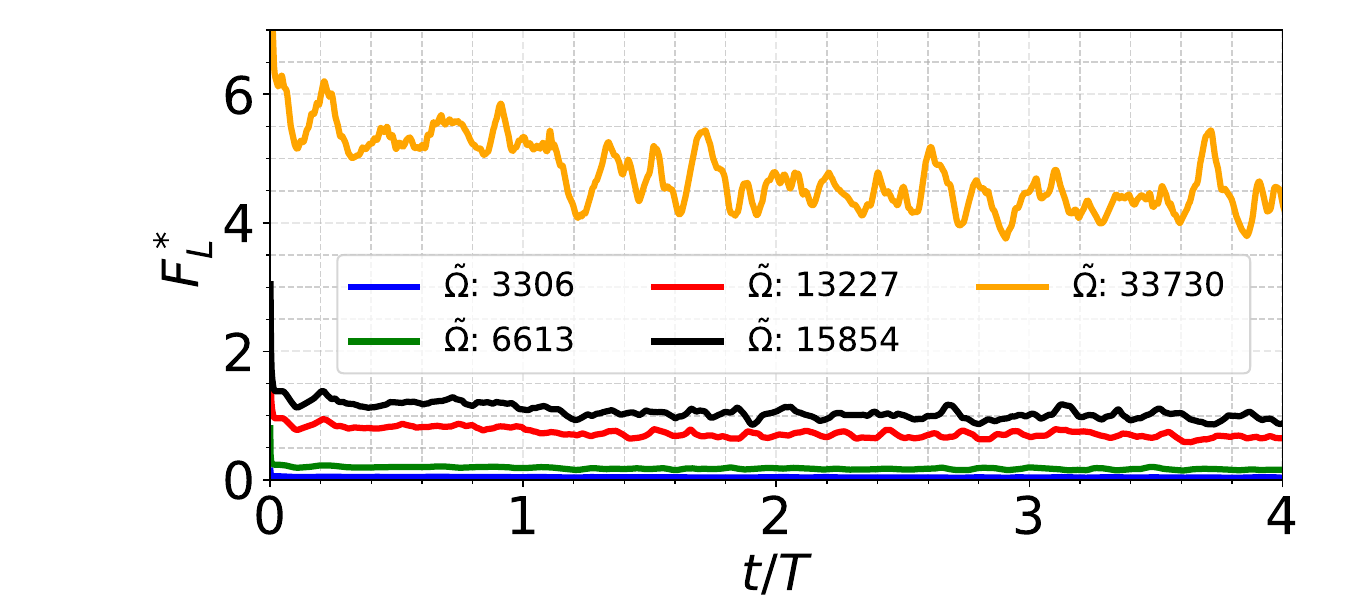}
         \caption{}
         \label{fig:45_Qcrit_all_case:e}
     \end{subfigure}
        \caption{The vortices shown using the iso-surface of the Q criterion for the RR-10 rotor at an pitch angle of 45$^\circ$ and corresponding to $\tilde{\Omega} =$  (a) 3307, (b) 6614, (c) 13228 and (d) 33730. The case with $\tilde{\Omega} =$ 15854 is presented in figure \ref{fig:45_Qcrit:a}. (e) The lift force shown for all the  $\tilde{\Omega}$ cases of the RR-10 rotor with pitch angle of 45$^\circ$ and normalized by the mean lift (between cycles 2 and 4) of the $\tilde{\Omega}$ = 15854 case.}
\label{fig:45_Qcrit_all_case}
\end{figure}

We begin by simulating the rectangular blade at a pitch angle of $45^o$ for a range of $\tilde{\Omega}$ values. Fig.  \ref{fig:45_Qcrit_all_case:a} - \ref{fig:45_Qcrit_all_case:d} show the vortex structures for 4 of these cases and figure \ref{fig:45_Qcrit_all_case:e} shows the time variation of the lift over 4 cycles of rotation for four cases with $\tilde{\Omega}$ ranging from 3,306 to 33,730. The domain size for these simulations were selected to be $10R_o \times 10R_o \times 8R_o$ in the x, y and z directions respectively. In the z-direction, the uniform fine mesh region was confined to $5c_h$ region (here, $c_h$ is the rotor chord length) placed such that $3c_h$ below and $2c_h$ region above the rotor center is resolved. The fine region in the plane of rotation (x-y plane) was $2.4R_o \times 2.4R_o$ box with the center of the box located at the center of the rotation and $1.1R_o \times 3c_h$ region surrounding the blade was uniform fine mesh and this was expanded very slowly away from the uniform region. The overall grid size ranged from 16 to 25 million grid points with more grid points used for larger area rotors. We conducted a grid refinement study, to ensure that the unsteady aerodynamic forces on the RR-10 blade were well converged on the current grid.  
We note that both the mean lift force as well as the fluctuations in the lift force increase with $\tilde{\Omega}$. The increase in mean lift will be discussed in detail in the following paragraph but the increase in fluctuation level is associated with the more intense unsteady vortex phenomena at higher $\tilde{\Omega}$ and this is expected to increase the broadband component of the noise emitted from the blade.

\begin{figure}
\centering
 \begin{minipage}{0.55\textwidth}
\includegraphics[width=1.\textwidth]{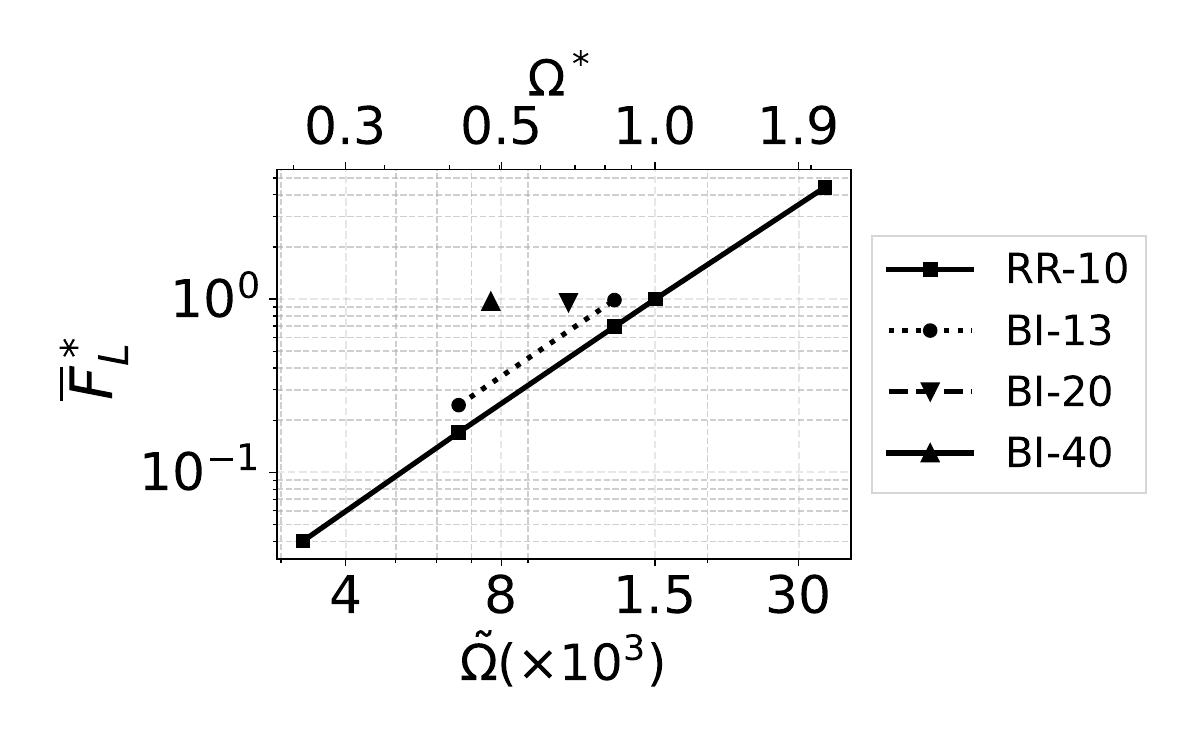}
\caption{Simulations for the four rotors cases with a pitch angle of 45$^\circ$ and with varying rotation speed to pick the cases with similar mean lift. The corresponding values are also presented in table \ref{table:comp_all_45_values}.}
\label{fig:lift:trials}
 \end{minipage}%
    \hfill
    \begin{minipage}{0.41\textwidth}
        \centering
        \begin{tabular}{|c|c|c|c|}
            \hline
            \textbf{Rotor} & \rbd{\textbf{$\tilde{\Omega}$}} & \rb{\textbf{$\bar{F}_L^*$}} & \rb{\textbf{$\bar{C}_L$}} \\
            \hline
            RR-10   & 3307 & 0.039 & 0.500\\
            \hline
            RR-10   & 6614 & 0.169 & 0.536\\
            \hline
            RR-10   & 13228 & 0.693 & 0.547\\
            \hline
            \textit{RR-10}   & 15854 & \textit{1.000} & 0.549\\
            \hline
            RR-10   & 33730 & 4.416 & 0.536\\
            \hline
            \hline
            BI-13   & 6614 & 0.244 & 0.580\\
            \hline
            BI-13   & 13228 & 0.987 & 0.585\\
            \hline
            BI-20   & 10782 & 0.957 & 0.568\\
            \hline
            BI-40   & 7631 & 0.969 & 0.575\\
            \hline
        \end{tabular}
        \captionof{table}{The non-dimensional rotation speed ($\tilde{\Omega}$), normalized lift and normalized lift coefficient shown for all the simulation with pitch angle of 45$^\circ$ and the baseline case used for normalization is indicated using italics.}
        \label{table:comp_all_45_values}
    \end{minipage}
\end{figure}

The mean values of lift are extracted from the last 2 cycles of the simulation, and Fig. \ref{fig:lift:trials} shows a plot of mean lift force normalized by the value corresponding to the $\tilde{\Omega}$=15,854 case, which we employ as the baseline case for the current study. Table \ref{table:comp_all_45_values} shows the mean values of this normalized lift $F^*_L$ and we note that the slope of the line on the log-log plot is 2.04, indicating a linear correlation between $\tilde{\Omega}^2$ and lift. This is confirmed by the lift coefficient which is varied between 0.500 and 0.549 for all the RR-10 cases simulated here. We note that while small-scale drone rotors have $\tilde{\Omega}$ values in the 200,000 range, such values are outside the capability of time-accurate simulations of the type used in the current study. However, current set of simulations shows that for sufficiently high values of $\tilde{\Omega}$, the lift coefficient becomes independent of this parameter. Thus, the results from the current study should be extendable to the higher $\tilde{\Omega}$ values relevant to practical small-scale drones.

We choose $\tilde{\Omega}$= 15854 as our baseline case for this part of the study and designate it as RR-10. Next, via trial and error assisted by estimates from the previously shown scaling laws, we determine a $\tilde{\Omega}$ for the other three blades that generate a similar mean lift as the baseline RR-10 case. These cases are designated as BI-13, BI-20, and BI-40 and their normalized lift and lift coefficients are shown in Table \ref{table:comp_all_45_values} and the figures. As expected, these blades with larger blade areas require a lower rotational speed to generate the same lift and the rotational speeds are 83\%, 68\%, and 48\% of the baseline case for BI-13, BI-20, and BI-40, respectively.

\begin{figure}
    \centering
 \begin{subfigure}[b]{0.24\textwidth}
      \centering         
      \includegraphics[width=\textwidth]{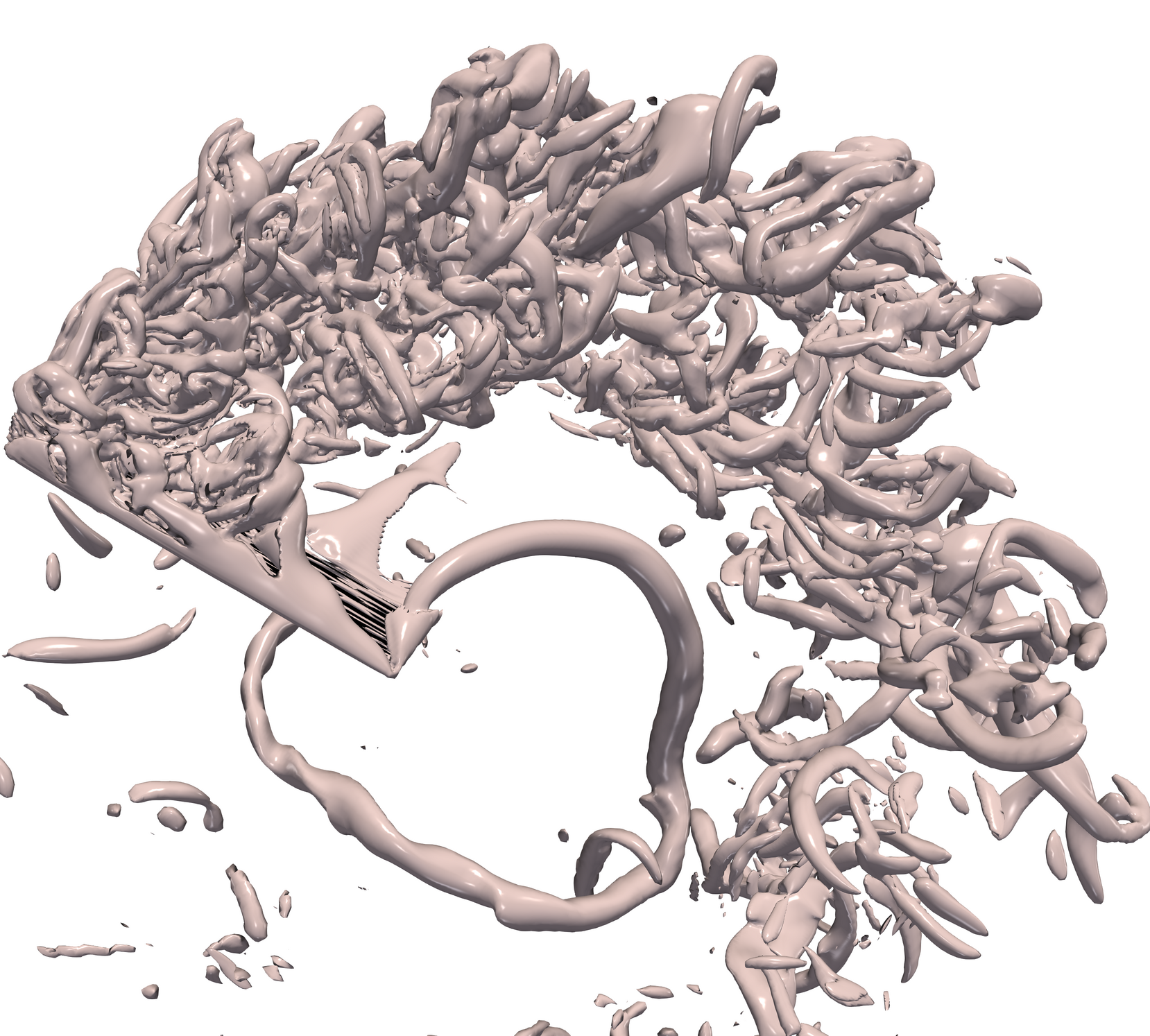}
         \caption{}
         \label{fig:45_Qcrit:a}
     \end{subfigure}
  \hfill
     \begin{subfigure}[b]{0.24\textwidth}
         \centering
         \includegraphics[width=\textwidth]{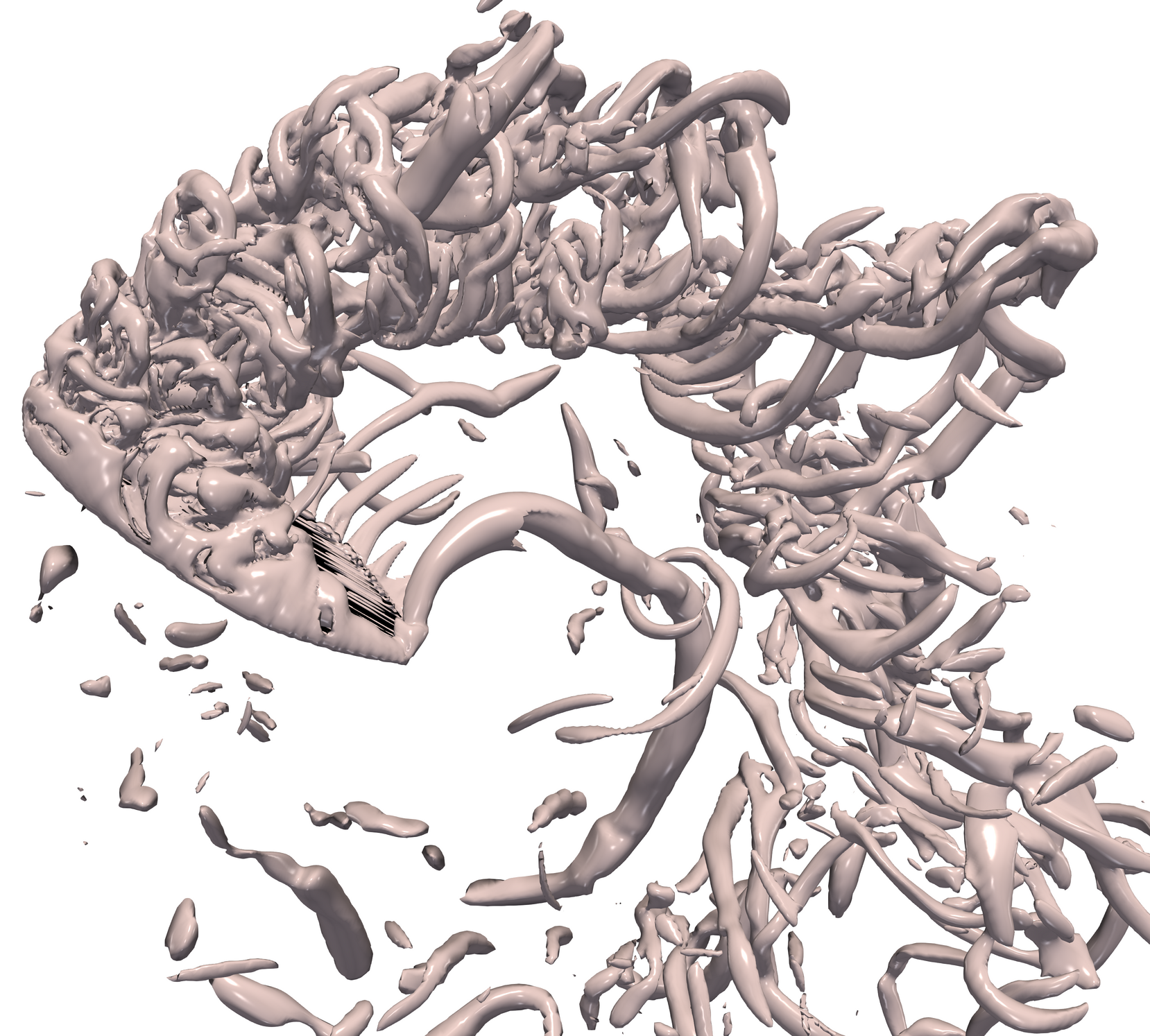}
         \caption{}
         \label{fig:45_Qcrit:b}
     \end{subfigure}
   \hfill
     \begin{subfigure}[b]{0.24\textwidth}
         \centering
         \includegraphics[width=\textwidth]{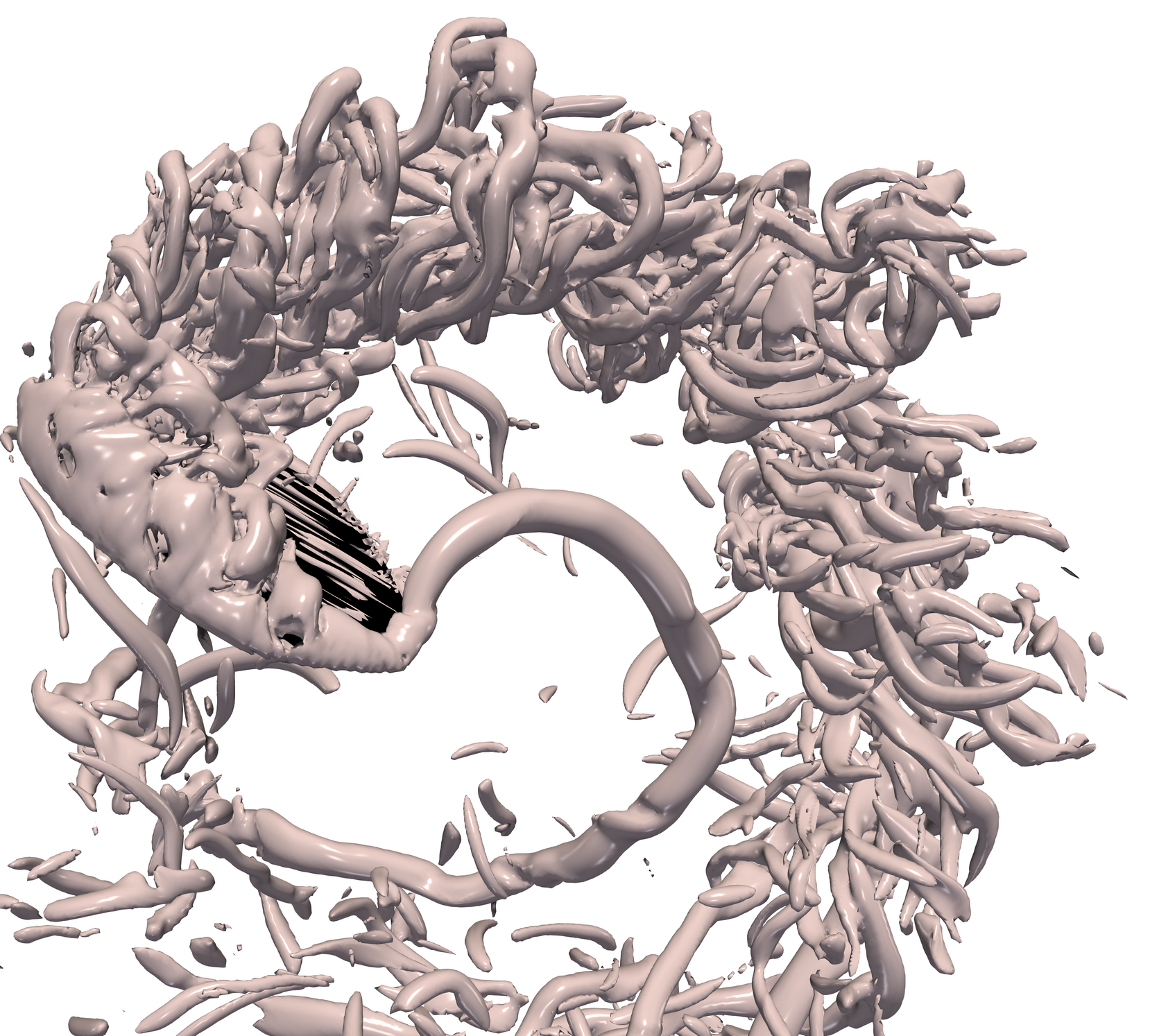}
         \caption{}
         \label{fig:45_Qcrit:c}
     \end{subfigure}
     \hfill
     \begin{subfigure}[b]{0.24\textwidth}
         \centering
         \includegraphics[width=\textwidth]{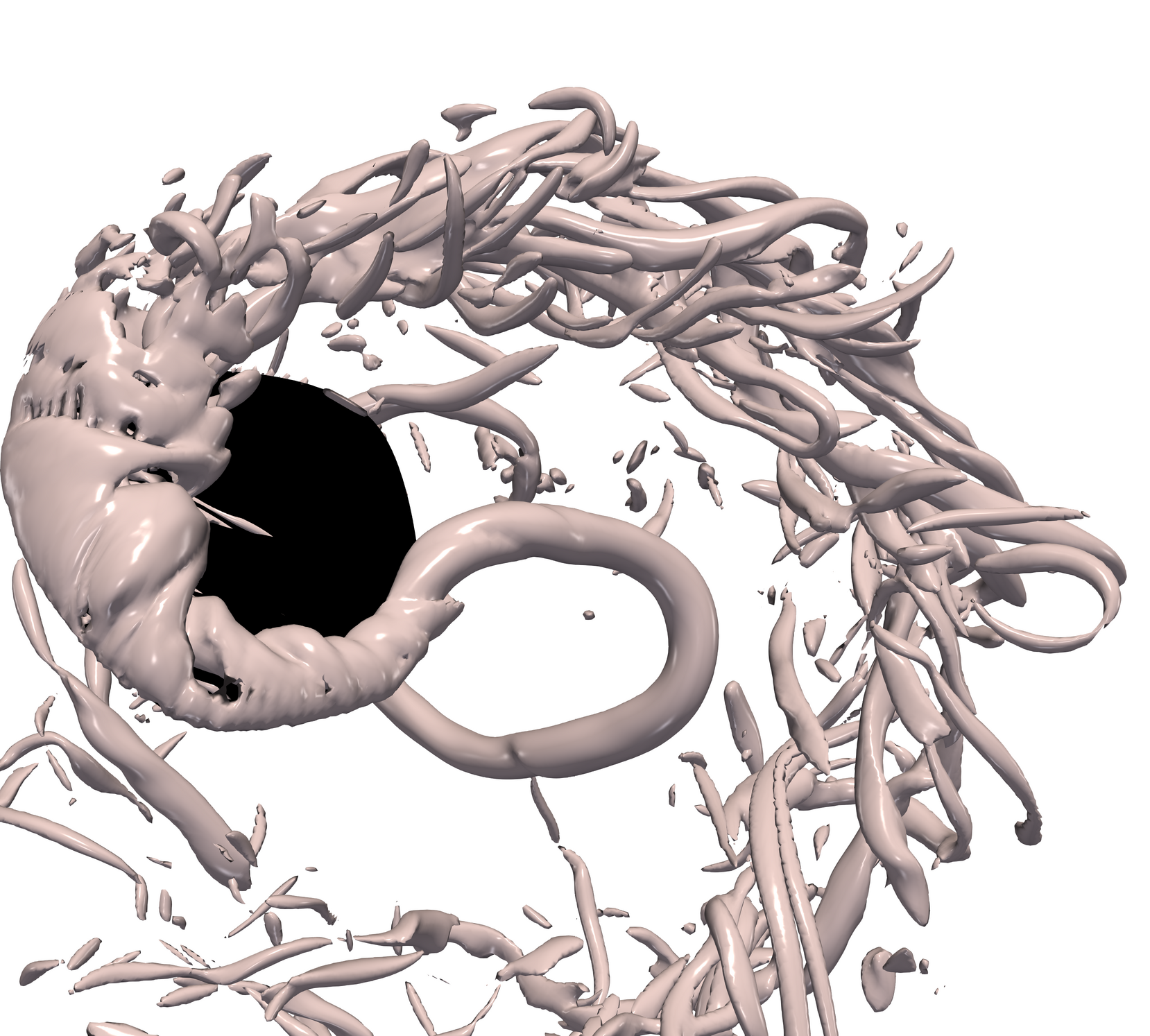}
         \caption{}
         \label{fig:45_Qcrit:d}
     \end{subfigure}
        \caption{The vortices corresponding to the pitch angle of 45$^\circ$ case shown for (a) rectangular rotor (RR-10) and bio-inspired rotors (b) BI-13, (c) BI-20 and (d) BI-40 using iso-surface of the Q criterion.}
\label{fig:45_Qcrit}
\end{figure}
The flow field corresponding to all four cases are shown in figure \ref{fig:45_Qcrit} with the vortical structures represented by the isosurface of the Q-criterion. We note that the blade with a higher aspect ratio such as the rectangular blade has two sets of counter-rotating vortices being shed from the leading edge while lower aspect ratio blades such as the elliptical blade of area four times the rectangular blade have only one rotating vortex near the leading edge. Also, the strength of the vortex is higher near the root and tip of the BI-40 blade whereas it is more evenly distributed for the rectangular rotor. 
These vortex structures were also observed by \cite{haibo_ar_study} in their study comparing wings with different aspect ratios and they attributed the presence of trailing edge vortices in higher aspect ratio blades to the satisfaction of the Kutta condition to balance out the stronger circulation of the leading edge vortex.
\begin{figure}
    \centering
 \begin{subfigure}[b]{0.48\textwidth}
      \centering
         \includegraphics[width=\textwidth]{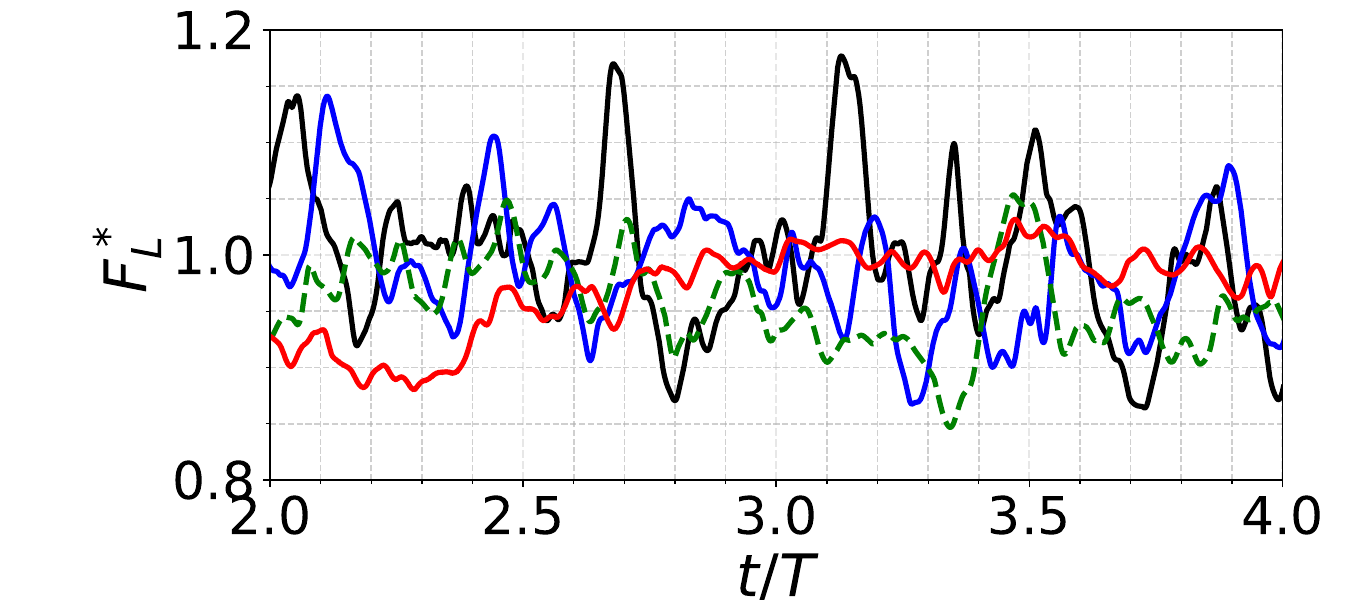}
         \caption{}
         \label{fig:LiftPowerComp:a}
     \end{subfigure}
  \hfill
     \begin{subfigure}[b]{0.48\textwidth}
         \centering
         \includegraphics[width=\textwidth]{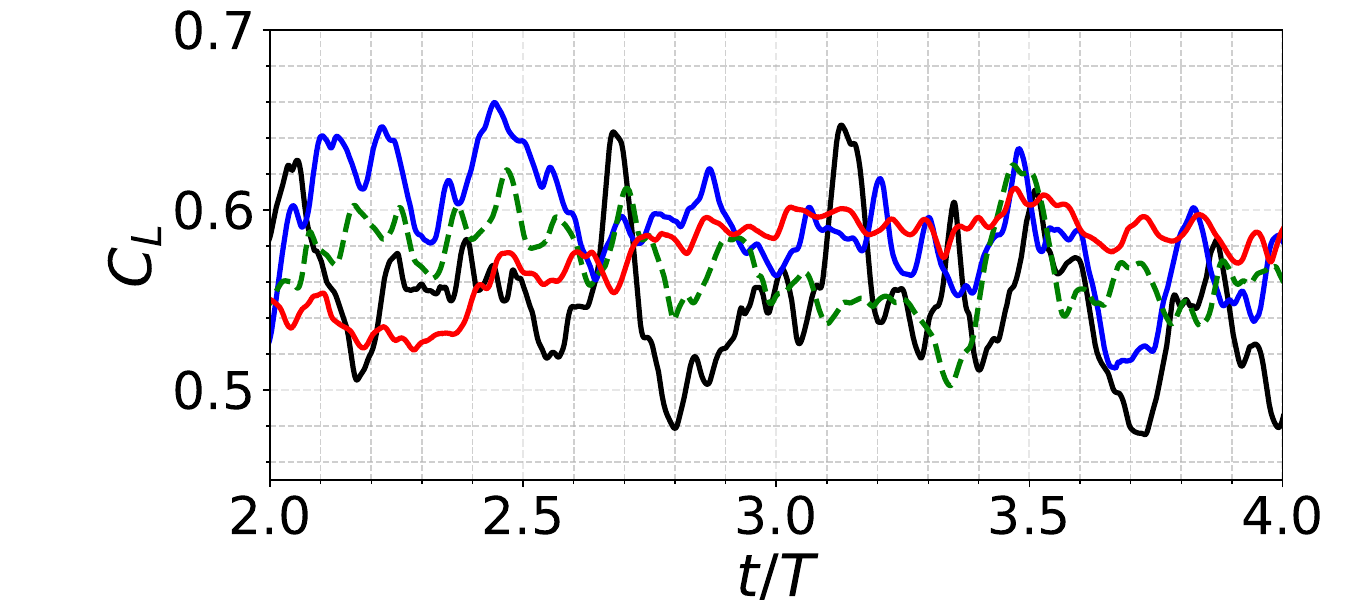}
         \caption{}
         \label{fig:LiftPowerComp:b}
     \end{subfigure}
   \hfill
     \begin{subfigure}[b]{0.48\textwidth}
         \centering
         \includegraphics[width=\textwidth]{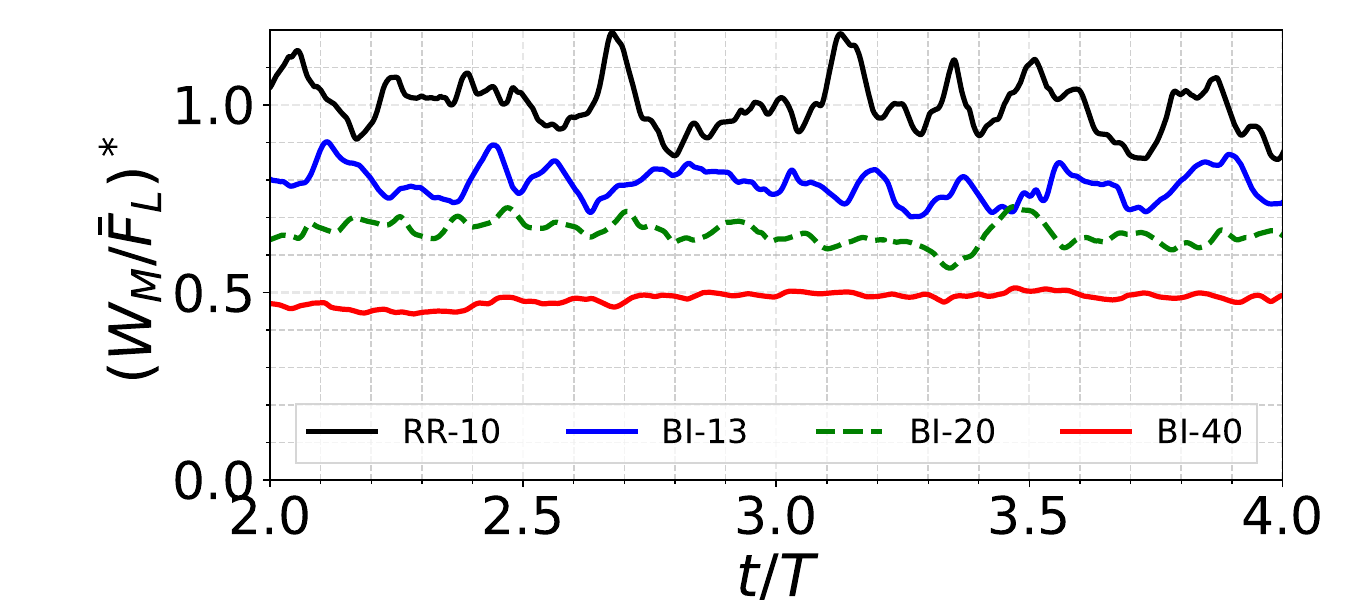}
         \caption{}
         \label{fig:LiftPowerComp:c}
     \end{subfigure}
     \hfill
     \begin{subfigure}[b]{0.48\textwidth}
         \centering
         \includegraphics[width=\textwidth]{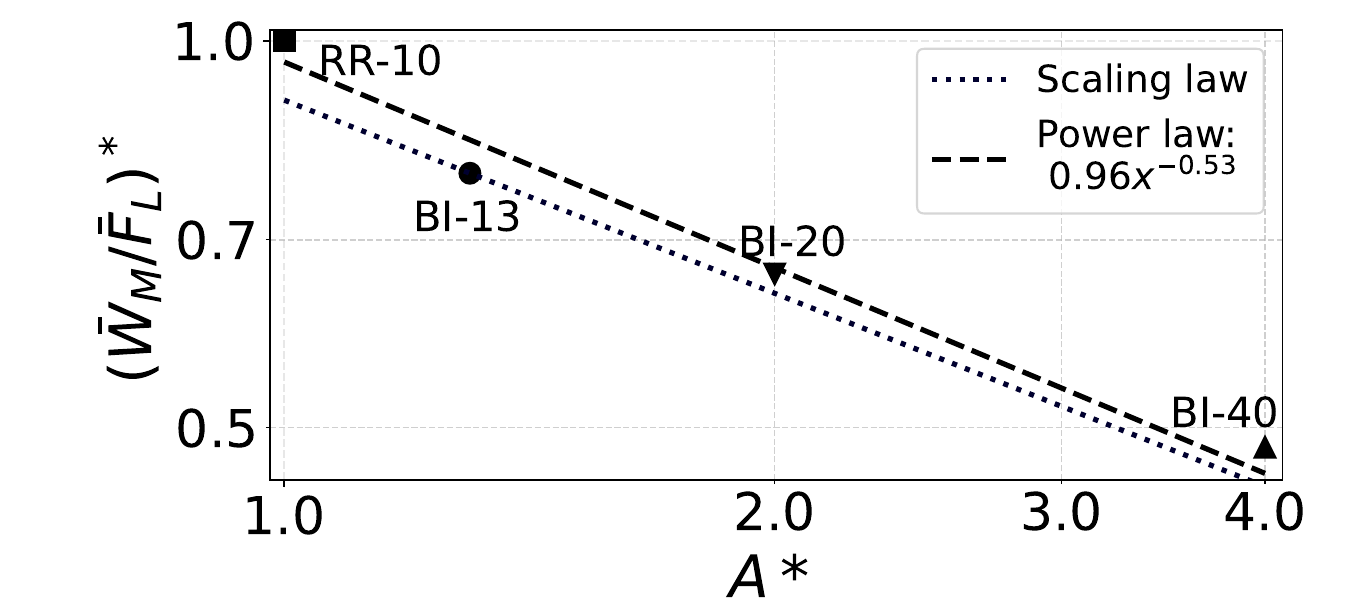}
         \caption{}
         \label{fig:LiftPowerComp:d}
     \end{subfigure}
        \caption{The (a) lift force, (b) lift coefficient based on the tip velocity and blade area, (c) mechanical power and (d) scaling of mechanical power with the area (equation \ref{eqn:power_scaling}) shown for 45$^\circ$ pitch angle rotors. Star superscript indicates that the values are normalized by the mean value for rectangular rotor.}
\label{fig:LiftPowerComp}
\end{figure}

The lift is shown in figure \ref{fig:LiftPowerComp:a} and we see that the mean lift is very similar for all these four cases with the difference between the largest and smallest mean lift being less than 3\%. We note that the fluctuation level in the lift is the largest for the RR-10 case and this is associated with the higher value of $\tilde{\Omega}$ for this blade. This higher level of force fluctuation will generate a higher intensity of broadband noise. Fig. \ref{fig:LiftPowerComp:b} shows the time variation of the lift coefficient for these four blades and the values for the four cases lie between about 0.475 and 0.66.

The aerodynamic power per unit lift for these rotors is shown in figure \ref{fig:LiftPowerComp:c} and Table \ref{table:45deg_rotor_comparision} includes the mean values of the key force, power, and acoustic metrics for these four blades.  We note that the rectangular rotor has the highest specific power requirement with the specific power decreasing as the area of the blade is increased from 1.3 to 4 times the area of the rectangular rotor. BI-40, which has four times the area of the rectangular RR-10 blade, has a specific power that is less than half the baseline rotor. 

The scaling of specific power with blade area is shown in figure \ref{fig:LiftPowerComp:d} and we see that the numerical simulation results are in good agreement with the scaling law of mechanical power with the area of the rotor shown in equation \ref{eqn:power_scaling}. This close match between the scaling law and the numerical simulation reflects the fact that the mean $C_L$ (and therefore the $C_D$) are maintained unchanged despite the significant changes in the shape of the blade.
\begin{figure}
    \centering
 \begin{subfigure}[b]{0.48\textwidth}
      \centering
         \includegraphics[width=\textwidth]{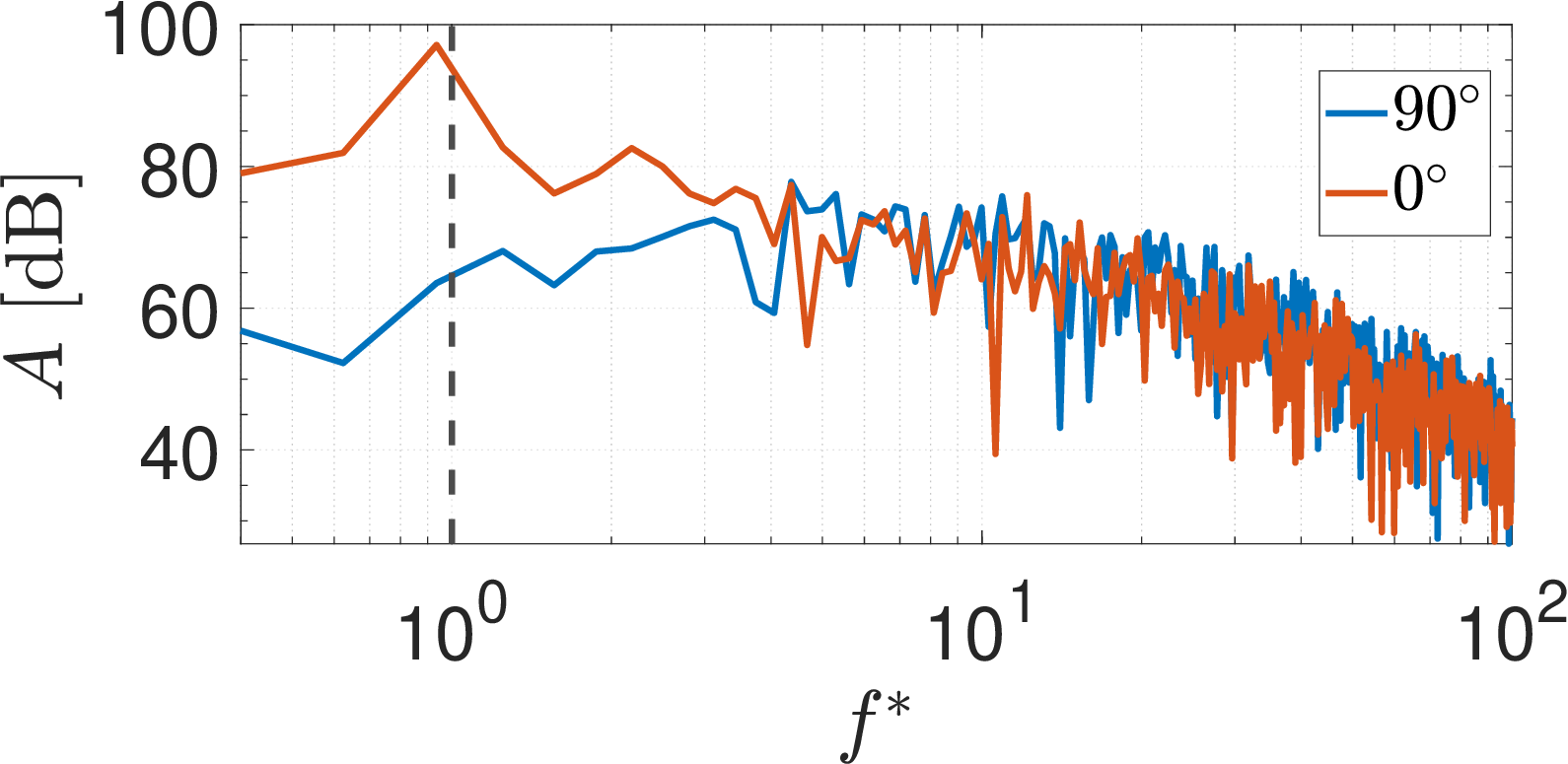}
         \caption{}
         \label{fig:directivity:a}
     \end{subfigure}
  \hfill
     \centering
 \begin{subfigure}[b]{0.48\textwidth}
      \centering
         \includegraphics[width=\textwidth]{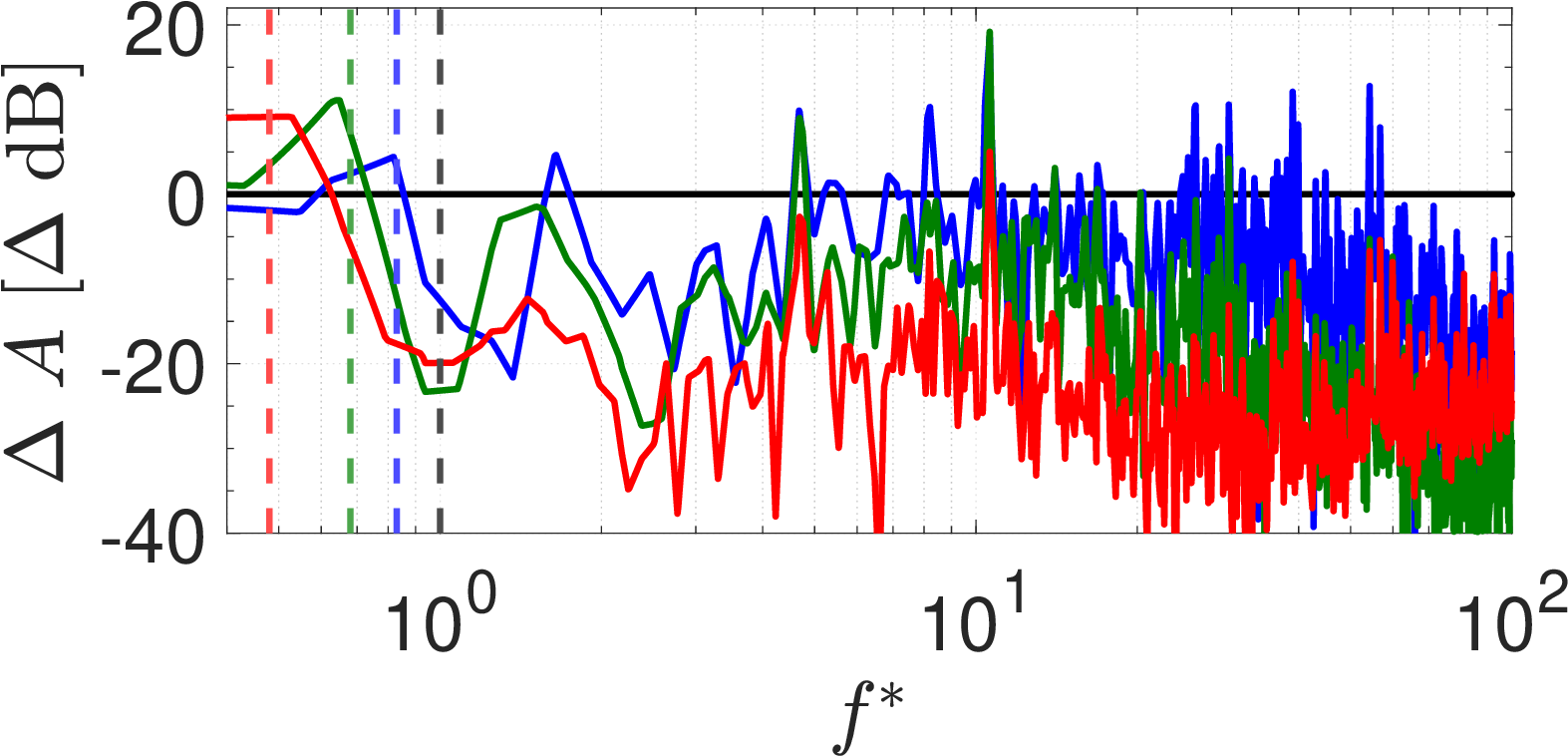}
         \caption{}
         \label{fig:directivity:b}
     \end{subfigure}
  \hfill
  \begin{subfigure}[b]{0.48\textwidth}
      \centering
         \includegraphics[width=\textwidth]{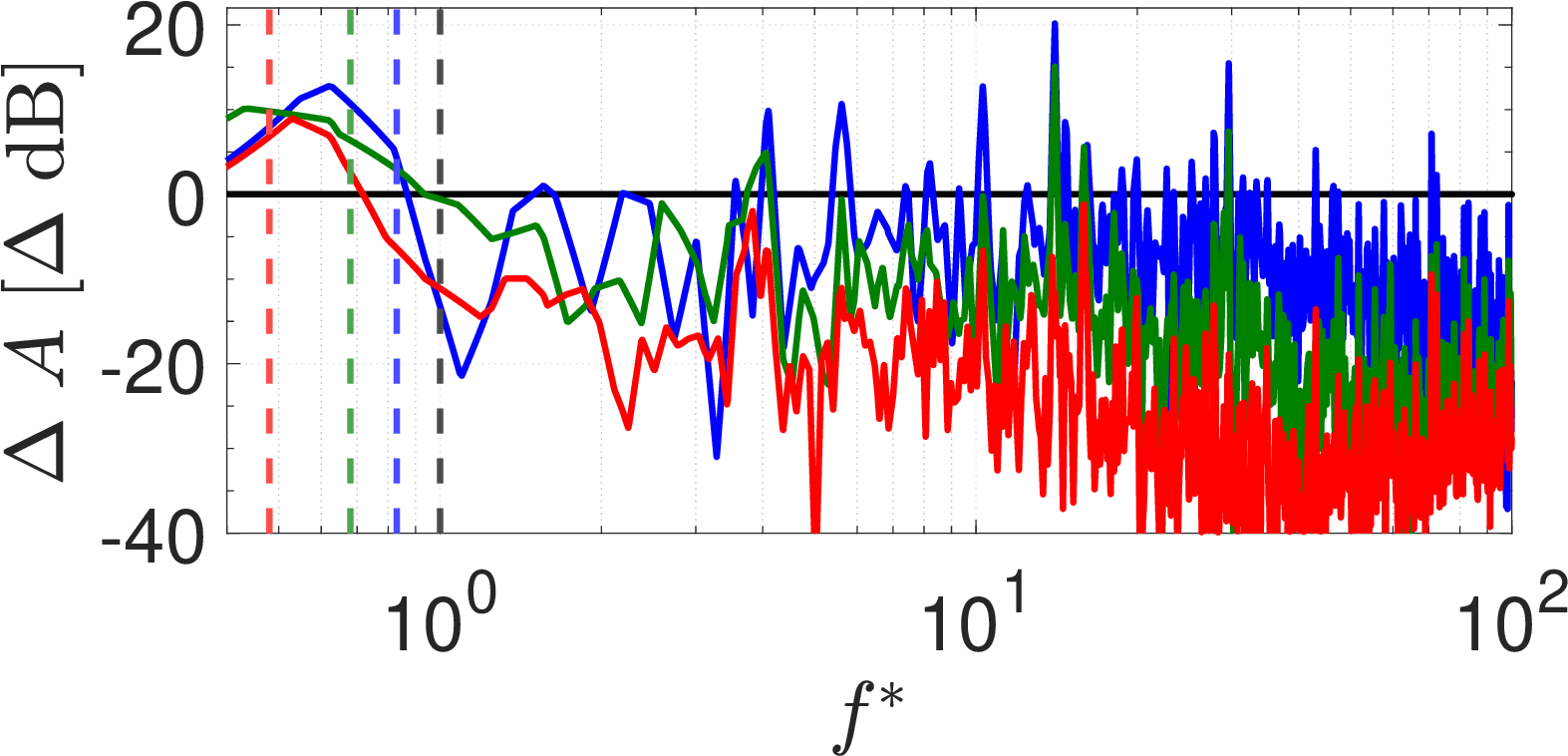}
         \caption{}
         \label{fig:directivity:c}
     \end{subfigure}
  \hfill
     \begin{subfigure}[b]{0.48\textwidth}
         \centering
\,\,\,\,\,\,\,\,\,\,\,\includegraphics[width=0.9\textwidth]{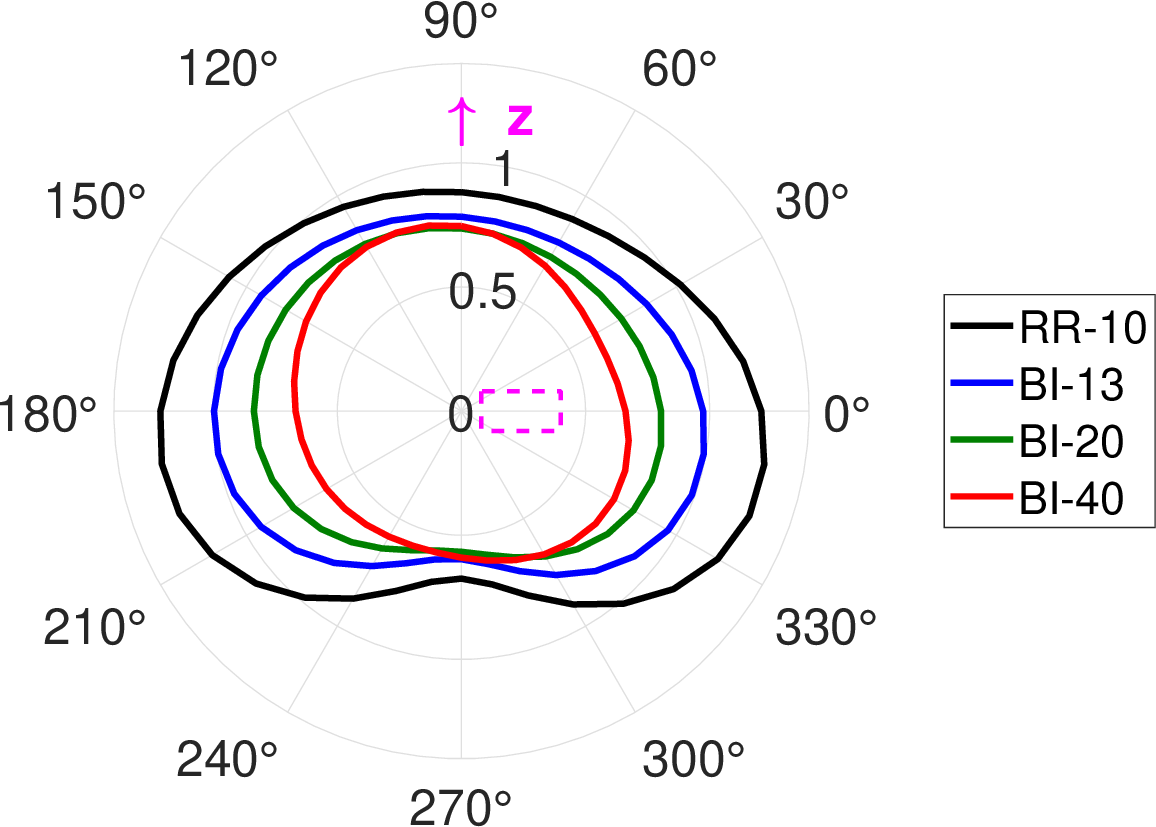}
         \caption{}
         \label{fig:directivity:d}
     \end{subfigure}
     \hfill
     \begin{subfigure}[b]{0.48\textwidth}
         \centering
         \includegraphics[width=\textwidth]{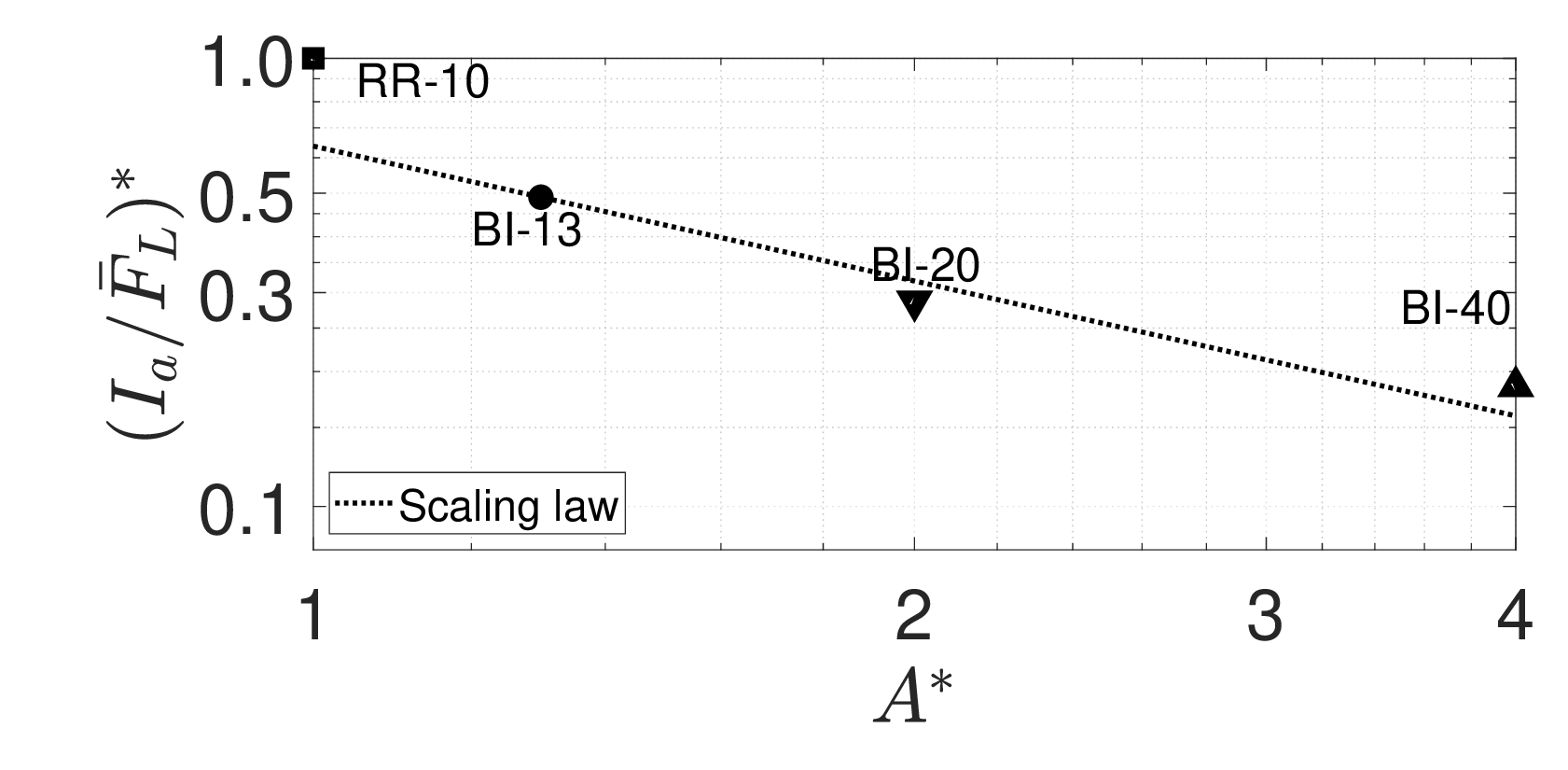}
         \caption{}
         \label{fig:directivity:e}
     \end{subfigure}
     \hfill
     \begin{subfigure}[b]{0.48\textwidth}
         \centering
         \includegraphics[width=\textwidth]{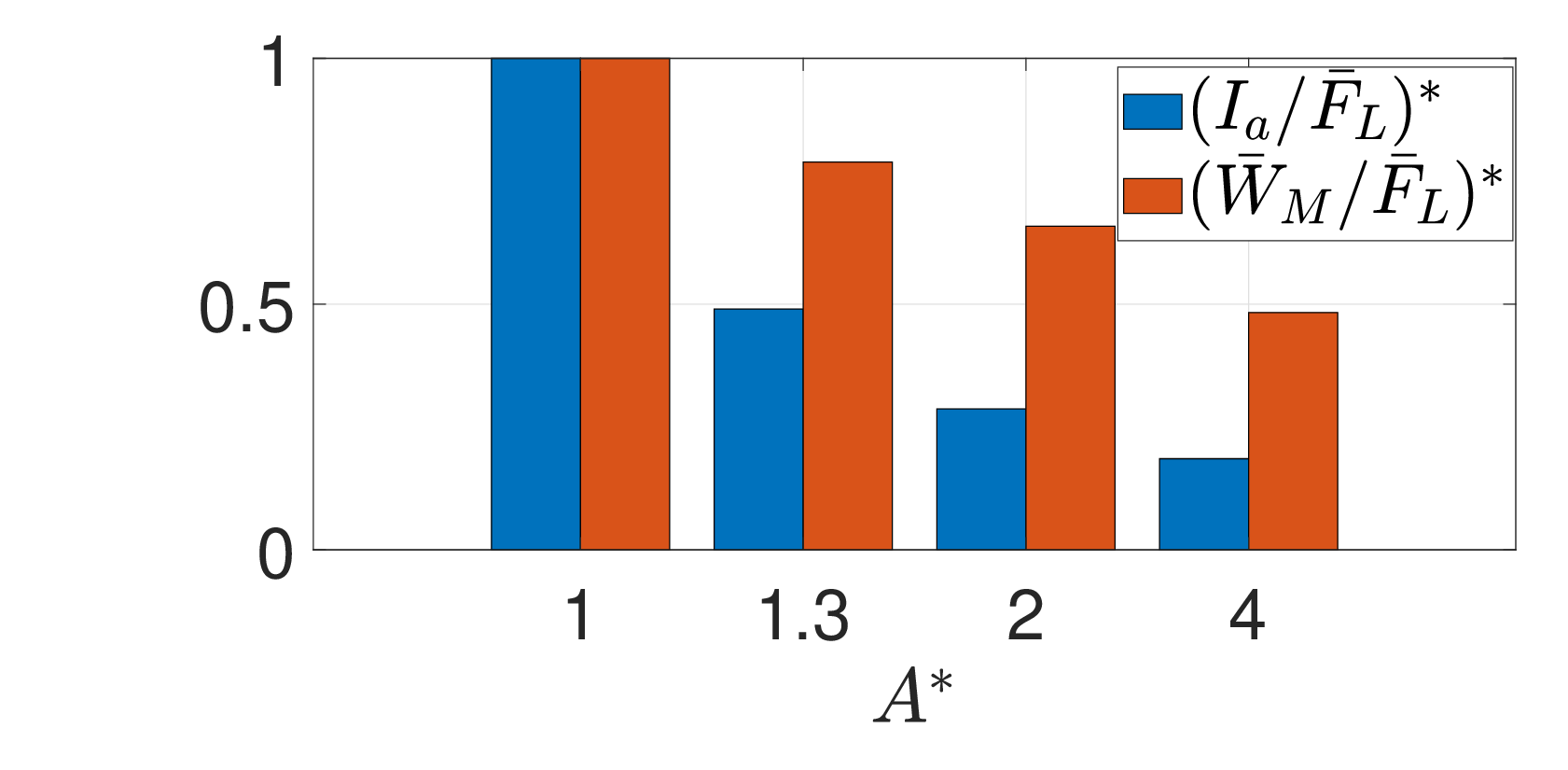}
         \caption{}
         \label{fig:directivity:f}
     \end{subfigure}
        \caption{Acoustic results shown for single blade rotors with 45$^\circ$ pitch angle. (a) The sound pressure level spectrum showing the amplitude of sound pressure ($A$) vs frequency ($f^*$) for the RR-10 rotor with frequency normalized using the rotation frequency of the RR-10 rotor. The sound spectrum is shown for all four rotors and calculated at a polar location of (b) (6.5$R_o$, 0$^\circ$) and (c) (6.5$R_o$, 90$^\circ$) with the frequency of the RR-10 rotor used to normalize the frequencies and the amplitude of the RR-10 rotor at each frequency is subtracted from the corresponding amplitude of all the rotors to show the gain or reduction in noise. \oc{The blade passage frequency is shown using dashed vertical lines.} d) The directivity pattern ($p_{rms}^{'*}$) at a distance of 6.5$R_o$ and normalized using the mean directivity corresponding to RR-10 case with dashed pink rectangle showing the blade orientation (not to scale), (e) the scaling of the acoustic intensity with blade area (equation \ref{eqn:intensity_scaling}) and (f) acoustic intensity and mean mechanical power shown for all four rotors. Star superscript indicates that the values are normalized by the corresponding value for rectangular rotor.}
\label{fig:directivity}
\end{figure}

The aerodynamic sounds are calculated at a monitoring distance of 6.5$R_o$ from the center of rotation for all the cases presented in this study. This distance between the rotor and the monitoring point is similar to the noise monitoring distance used in several drone aeroacoustics studies such as \cite{Hao_owl,noise_distance2,noise_distance3}.
We first show the sound pressure level spectrum for the RR-10 rotor in the figure \ref{fig:directivity:a} at two polar locations of (6.5$R_o$ , 0$^\circ$) and (6.5$R_o$ , 90$^\circ$). The angle $0^\circ$ lies in the plane of rotation and the spectrum shows a peak at the rotation frequency ($f^*=1$) associated with the rotation of the drag vector. The $90^\circ$ corresponds to a point along the vertical z-axis and since the noise in this direction depends only on the higher frequency fluctuation of the lift force, we do not see the blade tonal component at this location. The pitch angle of $45^\circ$ results in an equality of the lift and drag forces, and this results in the  broadband noise being similar for both these locations.

The sound spectrum for all four rotors is shown in figures \ref{fig:directivity:b} and \ref{fig:directivity:c} corresponding to polar locations of (6.5$R_o$ , 0$^\circ$) and (6.5$R_o$ , 90$^\circ$) respectively with the amplitude corresponding to RR-10 rotor subtracted from them to show the relative reduction or gain in the noise at each frequency. The larger area rotors rotate with lower frequency, and we observe the peak tonal noise being shifted to lower frequencies. We note that the rectangular blade which is operating at a higher $\tilde{\Omega}$ has larger broadband noise and as the area of the blades is increased, the broadband noise is also reduced with a reduction in $\tilde{\Omega}$. This is an additional benefit of increased blade area. 

The directivity pattern of the root-mean-squared sound pressure is shown in figure \ref{fig:directivity:d} and we note that while noise is reduced in all directions for BI blades, the reduction is particularly significant along planes that are aligned closer with the horizontal plane. This is due to the reduction in blade tonal noise which dominates the sound in this direction and is directly proportional to the rotation rate. In the lift direction, the noise depends on the fluctuations in the lift force and while these fluctuations are reduced as well for the larger area blades, the overall magnitude of this broadband noise is much smaller compared to the blade tonal ( $\sim$100 dB for the blade tonal versus $\sim$75 dB for the broadband noise - see Fig. \ref{fig:directivity:a}) and therefore this makes a smaller contribution to the overall sound and its reduction. The acoustic intensity per unit lift (i.e. specific acoustic power) for the four cases is shown in figure \ref{fig:directivity:e} and we see that the specific sound intensity decreases as the area of blades increases. However, this reduction does not strictly match the scaling analysis, especially between RR-10 and BI-13, where the reduction in sound intensity from the simulations exceeds that predicted by the scaling. This difference in the prediction and the scaling law is likely associated with the fact that the scaling law only accounts for the blade tonal where the total noise contains other components as well. 

The specific aerodynamic power and the specific sound intensity for all the rotors normalized by the value for RR-10 are shown in figure \ref{fig:directivity:f} with these results along with the mechanical power and the acoustic intensity normalized for the slightly different lift force produced by these rotors are summarized in table \ref{table:45deg_rotor_comparision}. This chart clearly shows that the reduction in rotation speed that is facilitated by the increase in blade area is indeed a promising strategy for reducing the noise and at the same time, improving the power efficiency of the rotor. We examine this further by varying some key features of the rotor configuration.
\begin{table}
\centering
\begin{tabular}{|l|c|c|c|c|}
\hline
\textbf{Quantity} & \textbf{RR-10} & \textbf{BI-13} & \textbf{BI-20} & \textbf{BI-40} \\
\hline
$\Omega^*$ & 1.00 & 0.834 & 0.680 & 0.481 \\
\hline
\rb{$\bar{F}_L^*$} & 1.00 & 0.987 & 0.957 & 0.969 \\
\hline
\rb{$\bar{C}_L$} & 0.549 & 0.585 & 0.568 & 0.575 \\
\hline
\rb{$\bar{W}_M^*$} & 1.00 & 0.778 & 0.630 & 0.468 \\
\hline
$I_a^*$ & 1.00 & 0.483 & 0.274 & 0.179 \\
\hline
\rb{$(\bar{W}_M/\bar{F}_L)^*$} & 1.00 & 0.788 & 0.658 & 0.482 \\
\hline
\rb{$(I_a/\bar{F}_L)^*$} & 1.00 & 0.489 & 0.286 & 0.185 \\
\hline
\end{tabular}
\caption{Summary of the mechanical power and aeroacoustic performance results for the pitch angle of 45$^\circ$ rotors compared to the baseline case of RR-10.}
\label{table:45deg_rotor_comparision}
\end{table}

\subsection{Blade Pitch \texorpdfstring{$\theta=20^\text{o}$}{theta20}}
As noted earlier, the lift coefficient depends on rotation rate, blade shape, and pitch angle. In the previous section, we have examined the effect of rotation rate and blade shape for a fixed pitch angle of 45$^\circ$, and in this section, we examine the effect of the pitch angle but repeating the study in the previous section to 20$^\circ$. In fact, drone rotors typically do not have pitch angles as high as 45$^\circ$ and this lower pitch angle is a more realistic representation of drone rotors. We therefore employ the same four rotors - RR-10 and BI-13, BI-20 and BI-40 but at 20 $^\circ$ pitch angle. As before, simulations are first performed to determine via trial-and-error, the rotation speeds of BI-13, BI-20, and BI-40 rotors that matched the lift to the RR-10 rotor. The normalized rotation speed ($\Omega^*$) corresponding to  RR-10 and BI-13, BI-20, and BI-40 are determined to be 1.0, 0.97, 0.88, and 0.71 respectively, for this pitch angle. 
\begin{figure}
    \centering
 \begin{subfigure}[b]{0.24\textwidth}
      \centering
         \includegraphics[width=\textwidth]{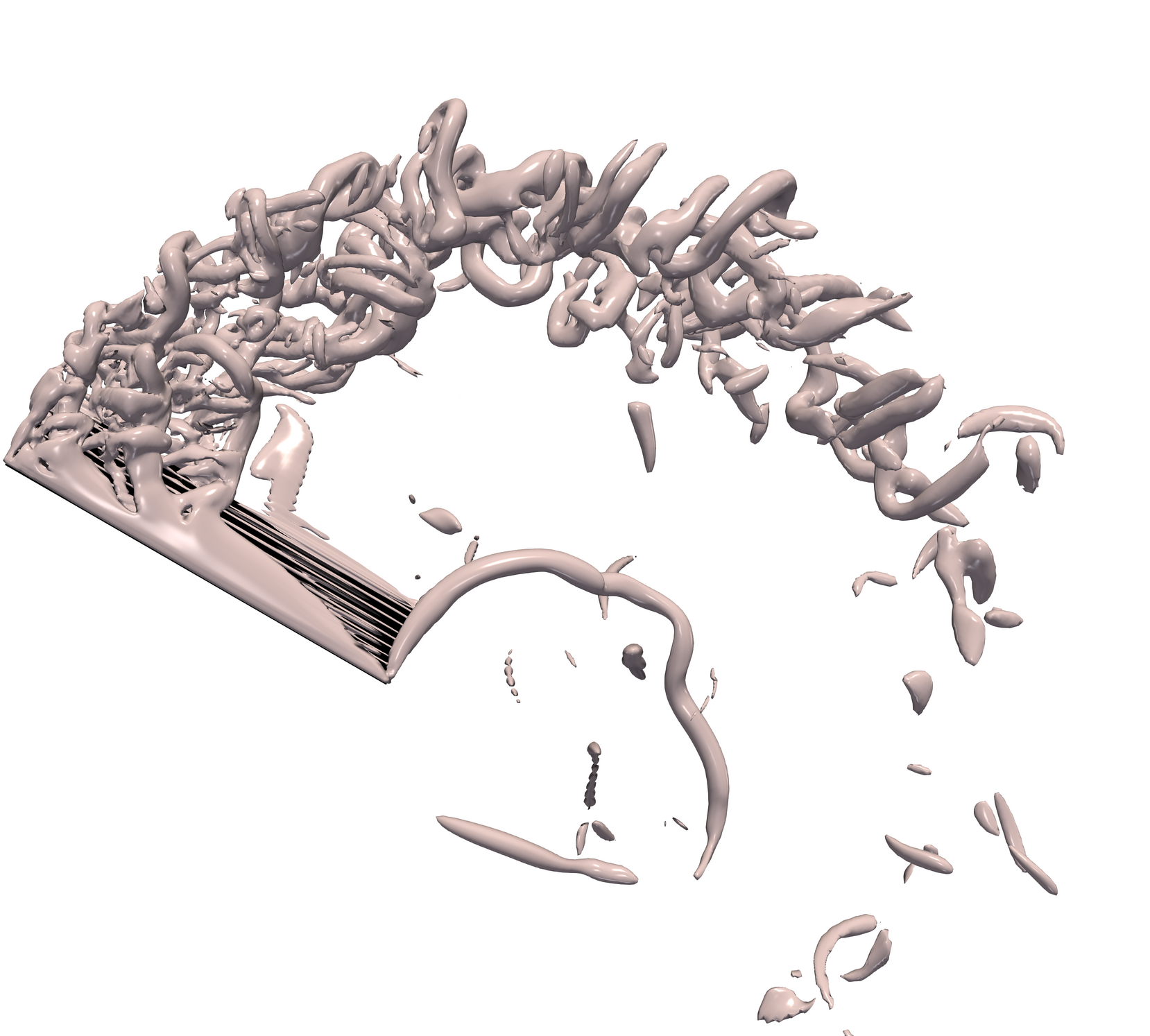}
         \caption{}
         \label{fig:20AoA_single:Qcrit:a}
     \end{subfigure}
  \hfill
     \begin{subfigure}[b]{0.24\textwidth}
         \centering
         \includegraphics[width=\textwidth]{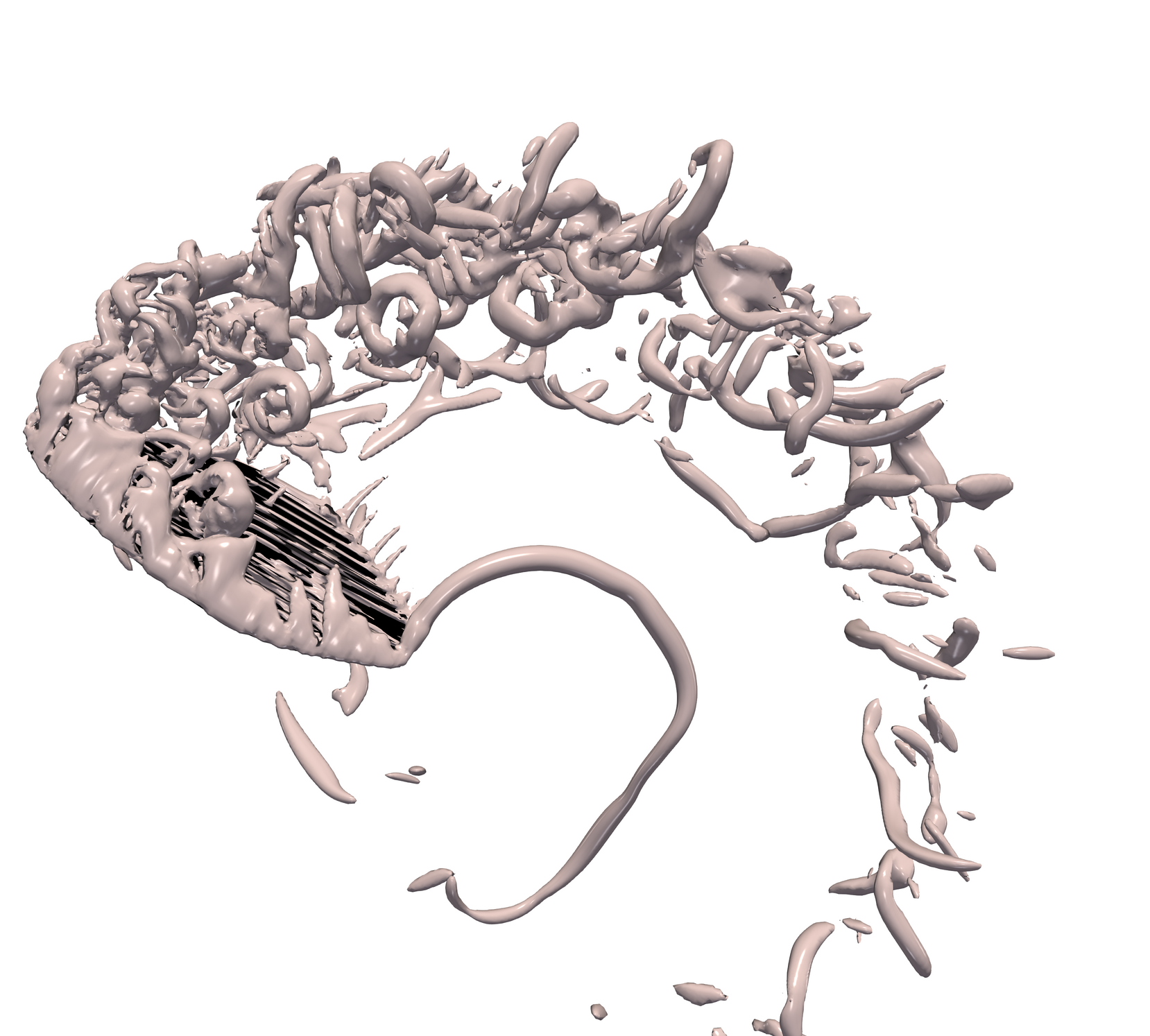}
         \caption{}
         \label{fig:20AoA_single:Qcrit:b}
     \end{subfigure}
   \hfill
     \begin{subfigure}[b]{0.24\textwidth}
         \centering
         \includegraphics[width=\textwidth]{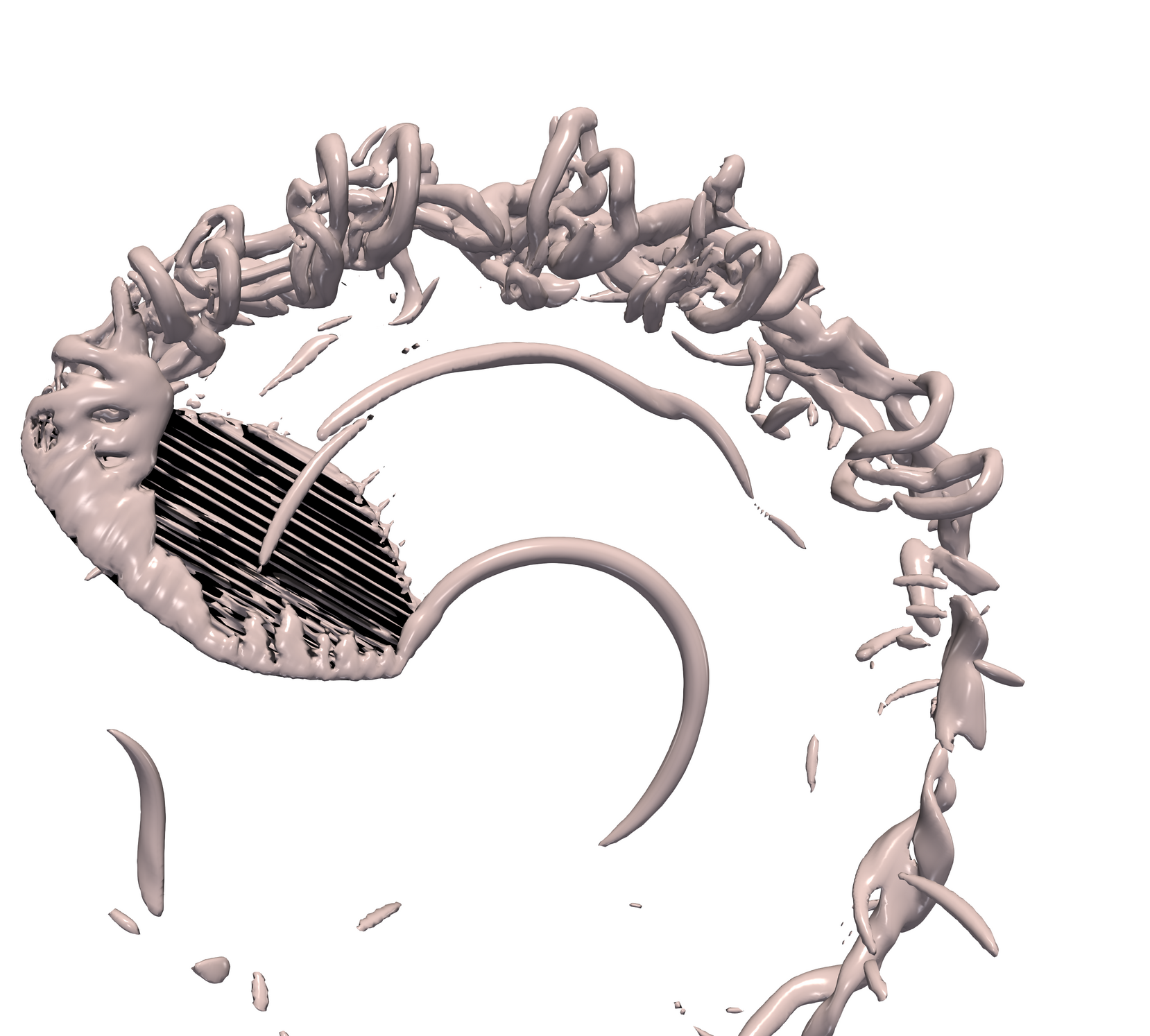}
         \caption{}
         \label{fig:20AoA_single:Qcrit:c}
     \end{subfigure}
     \hfill
     \begin{subfigure}[b]{0.24\textwidth}
         \centering
         \includegraphics[width=\textwidth]{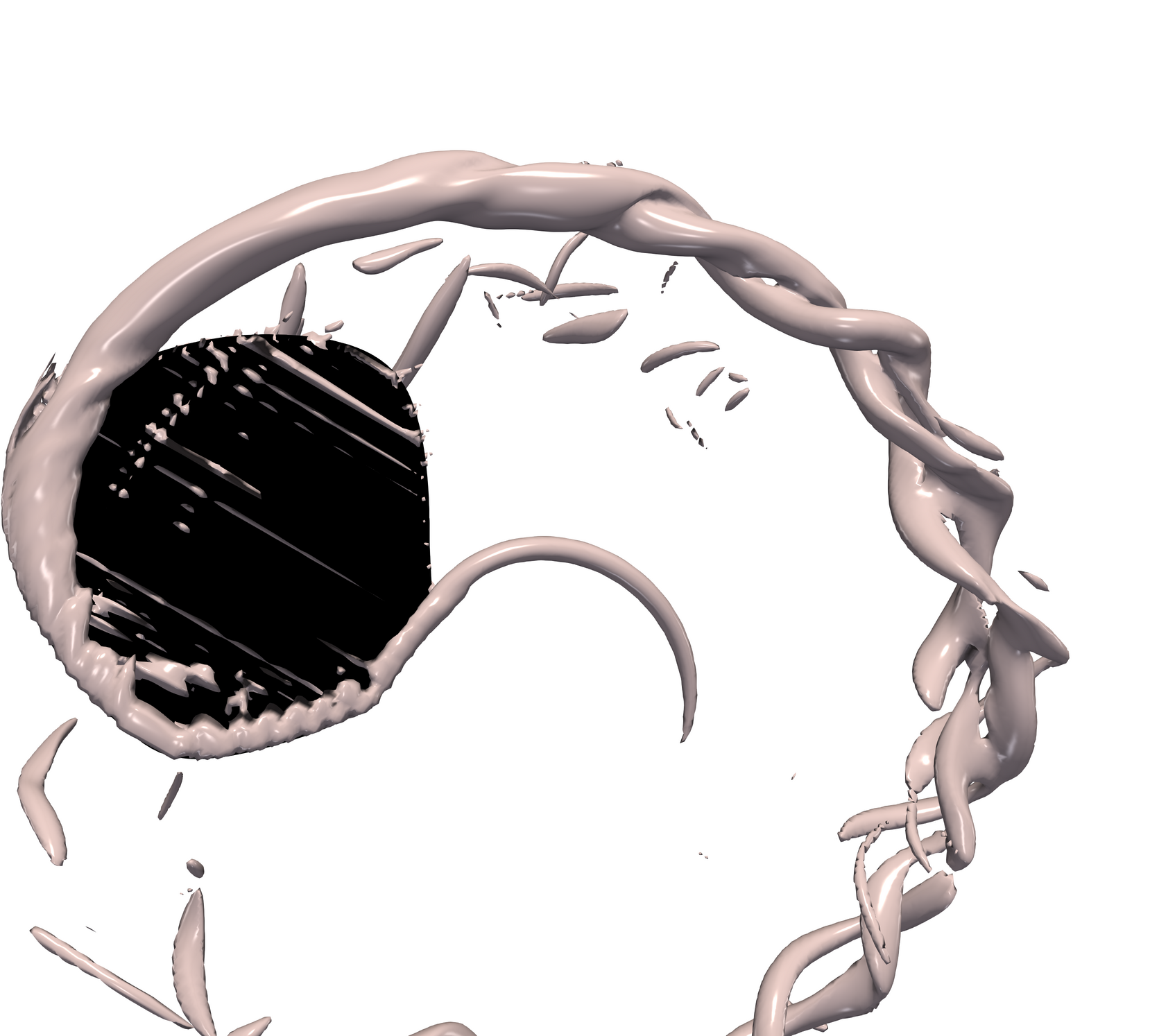}
         \caption{}
         \label{fig:20AoA_single:Qcrit:d}
     \end{subfigure}
        \caption{The vortices corresponding to the 20$^\circ$ pitch angle (a) rectangular rotor (RR-10) and bio-inspired rotors- (b) BI-13, (c) BI-20 and (d) BI-40 shown using the iso-surface of the Q criterion.}
\label{fig:20AoA_single:Qcrit}
\end{figure}
The vortex structures for these rotors are shown in figure \ref{fig:20AoA_single:Qcrit}. We note that the vortices shed from the rectangular rotor are concentrated near the outer half part of the span in a narrower region compared to the corresponding 45$^\circ$ pitch case. We also note a pair of counter-rotating vortices being formed and shed from the leading edge in the higher-aspect ratio RR-10 and BI-13 blade but only a single rotating tip vortex is seen for the BI-20 and BI-40 cases.

\begin{figure}
    \centering
 \begin{subfigure}[b]{0.48\textwidth}
      \centering
         \includegraphics[width=\textwidth]{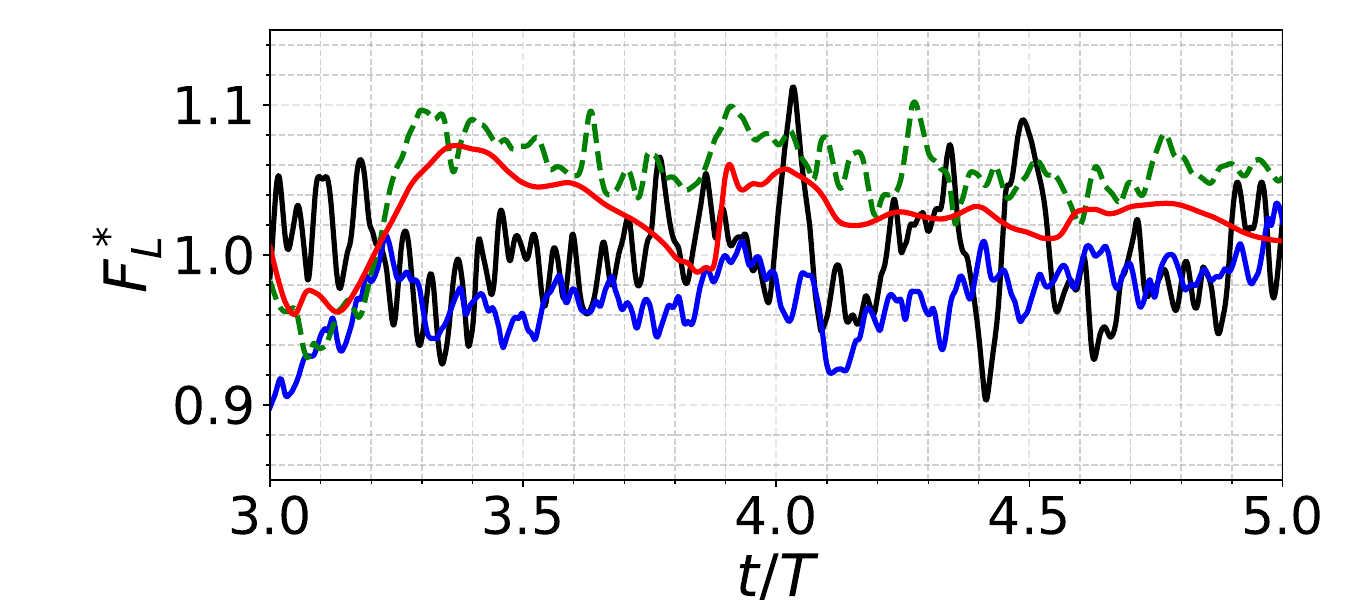}
         \caption{}
         \label{fig:aoa20:draglift:a}
     \end{subfigure}
  \hfill
     \begin{subfigure}[b]{0.48\textwidth}
         \centering
         \includegraphics[width=\textwidth]{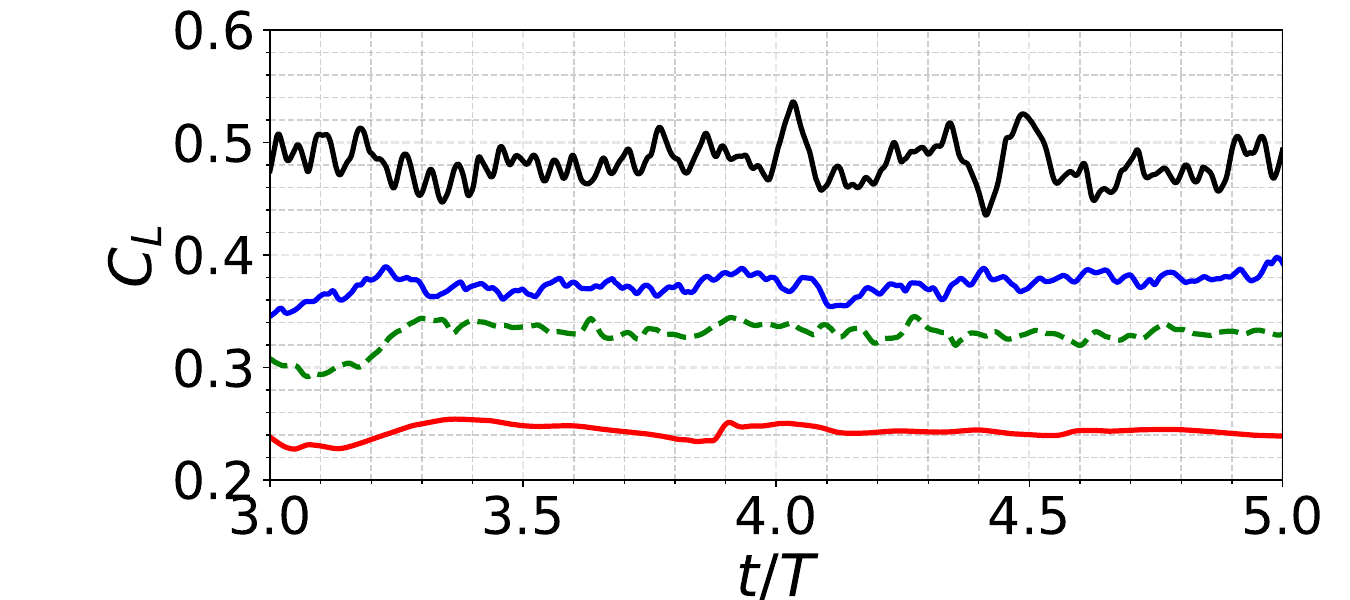}
         \caption{}
         \label{fig:aoa20:draglift:b}
     \end{subfigure}
   \hfill
     \begin{subfigure}[b]{0.48\textwidth}
         \centering
         \includegraphics[width=\textwidth]{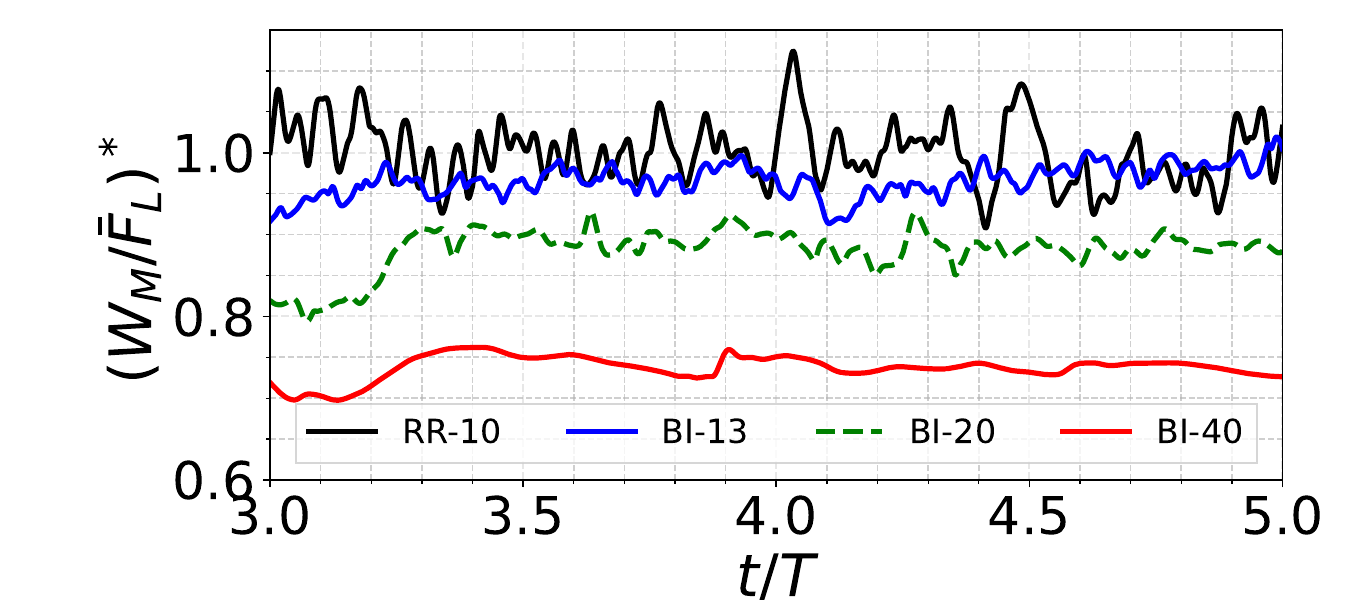}
         \caption{}
         \label{fig:aoa20:draglift:c}
     \end{subfigure}
     \hfill
     \begin{subfigure}[b]{0.48\textwidth}
         \centering
         \includegraphics[width=\textwidth]{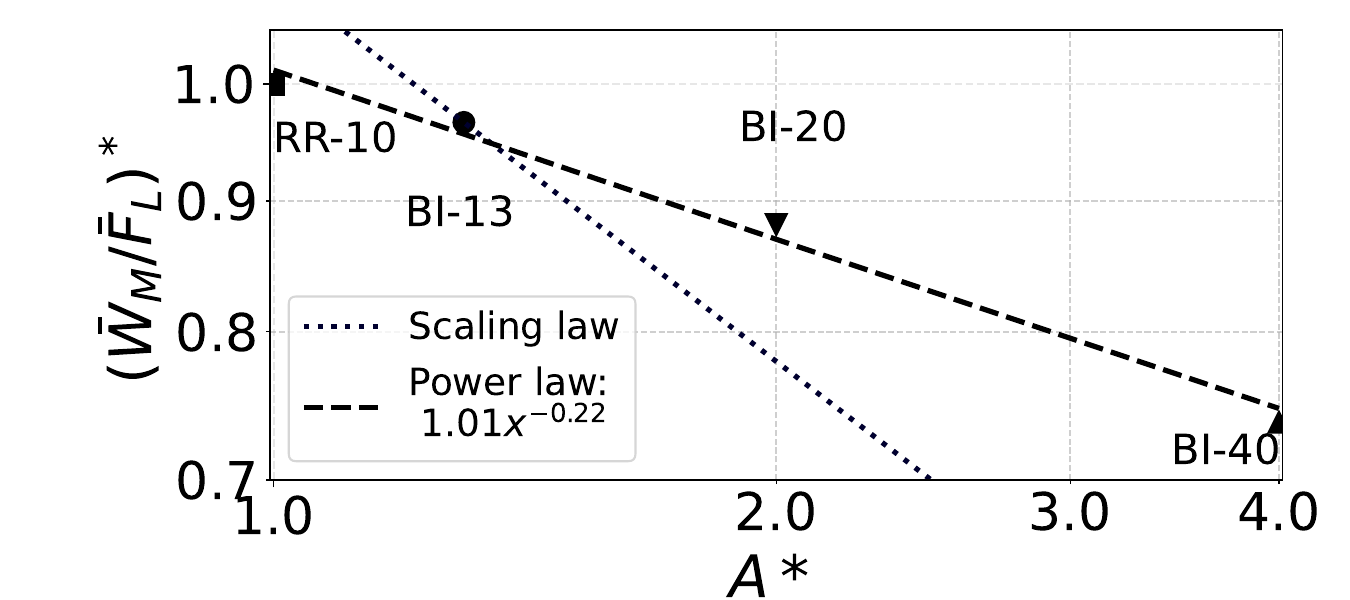}
         \caption{}
         \label{fig:aoa20:draglift:d}
     \end{subfigure}
        \caption{The (a) lift force, (b) lift coefficient normalized based on the tip velocity and blade area, (c) mechanical power and (d) scaling of mechanical power with area (equation \ref{eqn:power_scaling}) shown for the 20$^\circ$ pitch angle rotors. Star superscript indicates the corresponding quantities are normalized using mean value of RR-10.}
\label{fig:aoa20:draglift}
\end{figure}

The corresponding lift force for all four cases is shown in figure \ref{fig:aoa20:draglift:a} with the maximum difference between mean lift across these rotors being around 7.46\% (the mean in this section is calculated between cycles,$t/T =$ 3 to 5). The drag forces follow a similar pattern as the lift force and are not shown separately since the ratio of lift to drag for all these cases is determined by the pitch angle with the ratio for this case being 2.747. The lift coefficient for all these cases is shown in figure \ref{fig:aoa20:draglift:b} and we see that the rectangular rotor has the largest lift coefficient followed by the bio-inspired (BI) rotors with the lift coefficient decreasing as the area of the rotors is increased. Interestingly, the lift coefficient decreases significantly with increasing blade area in this case. This is likely due to the fact that at the large pitch angle, the flow over the blade is highly separated and less sensitive to the blade shape but for the lower pitch angle, the flow and the associated aerodynamic forces are more sensitive to the shape characteristics of the blade. The reduction in the lift coefficient with increasing blade area is expected to diminish the magnitude of reduction in the specific acoustic noise and specific aerodynamic power and we examine this through the subsequent plots.

The time variation of the specific aerodynamic power required for rotation for all these blades is shown in figure \ref{fig:aoa20:draglift:c} and the mean normalized values of this quantity alongside the scaling for the mean mechanical power with area are shown in figure \ref{fig:aoa20:draglift:d}. Table \ref{table:20deg_rotor_comparision} shows the mean values of these and other relevant quantities. The plot indicates that the specific aerodynamic power reduces monotonically with increased blade area. A best-fit line through the semi-log plot indicates that the reduction in specific power scales with $A^{-0.22}$ which is a slower decrease relative to  $A^{-0.50}$ that is expected from the scaling in Eq. \ref{eqn:power_scaling}. This deviation is associated with the reduction in the lift coefficient with blade area for this pitch angle, which requires the larger-area blades to rotate at higher speeds to maintain lift. However, despite this deterioration in performance, the reductions in specific power are substantial with the BI-40 blade requiring a specific power that is 74\% of RR-10.
\begin{table}
\centering
\begin{tabular}{|l|c|c|c|c|}
\hline
\textbf{Quantity} & \textbf{RR-10} & \textbf{BI-13} & \textbf{BI-20} & \textbf{BI-40} \\
\hline
$\Omega^*$ & 1.00 & 0.970 & 0.876 & 0.713 \\
\hline
\rb{$\bar{F}_L^*$} & 1.00 & 0.971 & 1.050 & 1.026 \\
\hline
\rb{$\bar{C}_L$} & 0.482 & 0.374 & 0.329 & 0.243 \\
\hline
\rb{$\bar{W}_M^*$} & 1.00 & 0.938 & 0.912 & 0.763 \\
\hline
$I_a^*$ & 1.00 & 0.571 & 0.598 & 0.461 \\
\hline
\rb{$(\bar{W}_M/\bar{F}_L)^*$} & 1.00 & 0.966 & 0.868 & 0.743 \\
\hline
\rb{$(I_a/\bar{F}_L)^*$} & 1.00 & 0.588 & 0.570 & 0.449 \\
\hline
\end{tabular}
\caption{Summary of the mechanical power and aeroacoustic performance results for the 20$^\circ$ pitch angle rotors and normalized using the baseline case of RR-10 rotor.}
\label{table:20deg_rotor_comparision}
\end{table}

\begin{figure}
    \centering
     \begin{subfigure}[b]{0.48\textwidth}
         \centering
         \includegraphics[width=\textwidth]{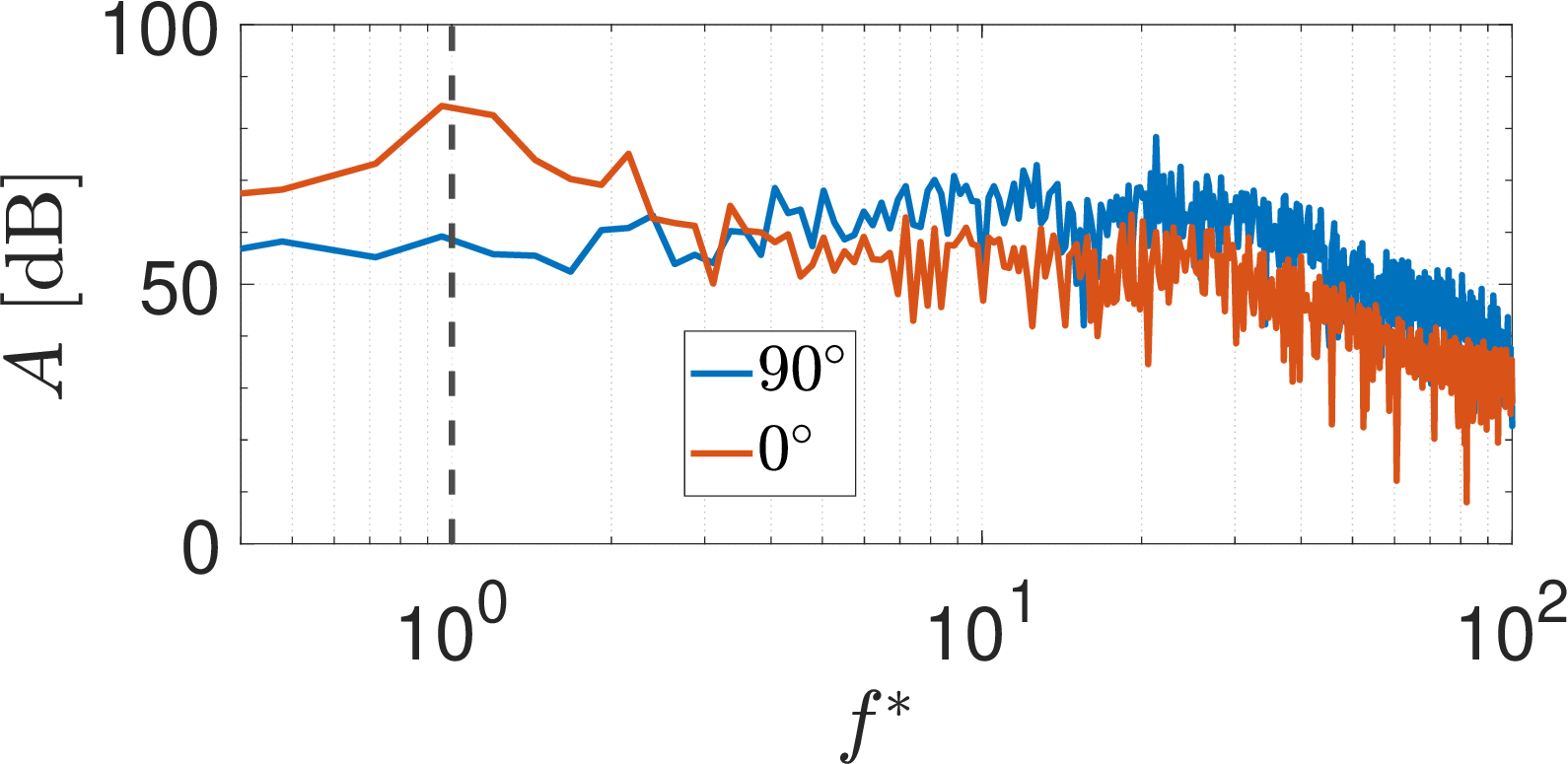}
         \caption{}
         \label{fig:aoa20:fwh:a}
     \end{subfigure}
     \hfill
     \centering
     \begin{subfigure}[b]{0.48\textwidth}
         \centering
         \includegraphics[width=\textwidth]{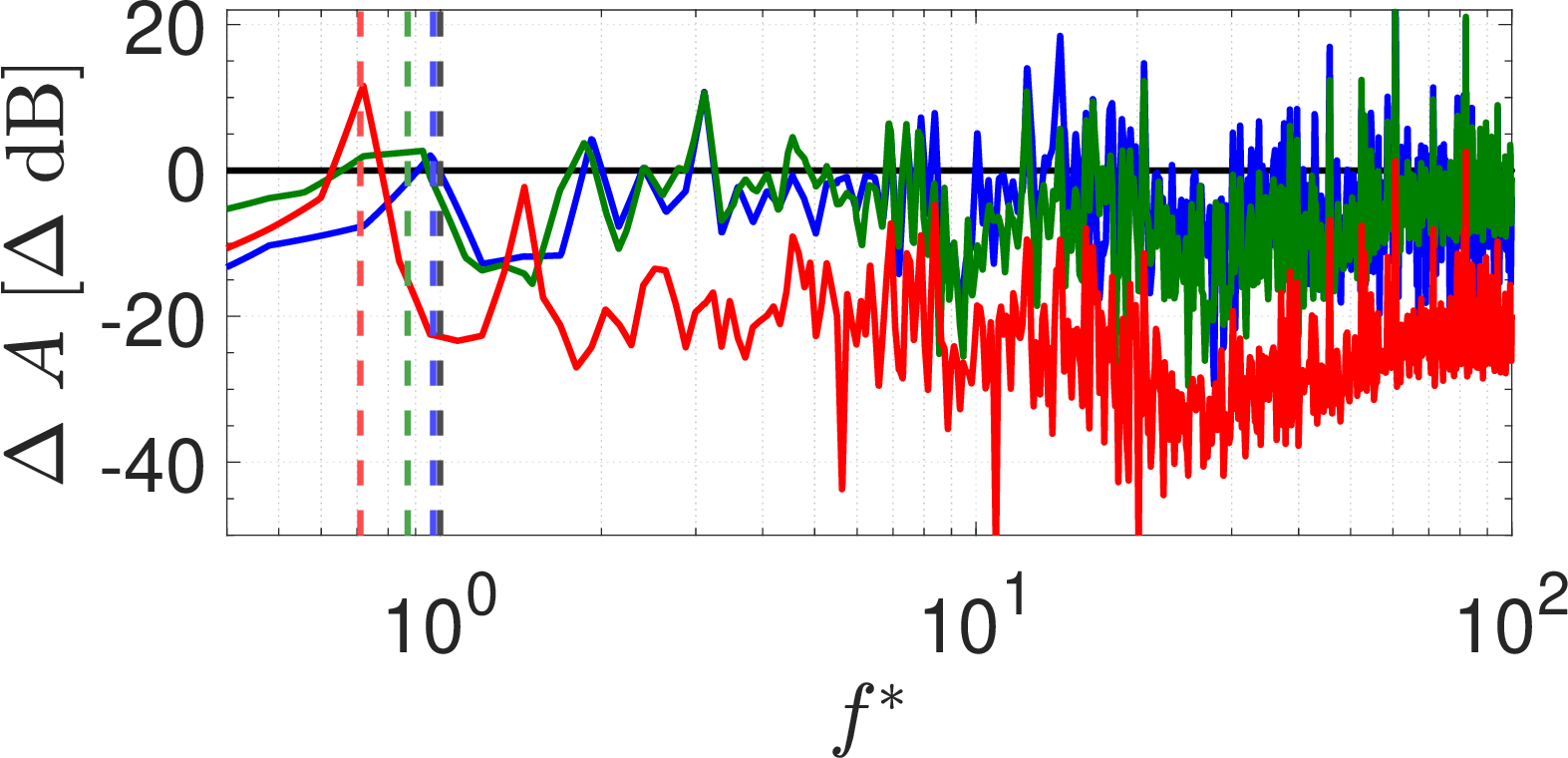}
         \caption{}
         \label{fig:aoa20:fwh:b}
     \end{subfigure}
     \hfill
     \centering
     \begin{subfigure}[b]{0.48\textwidth}
         \centering
         \includegraphics[width=\textwidth]{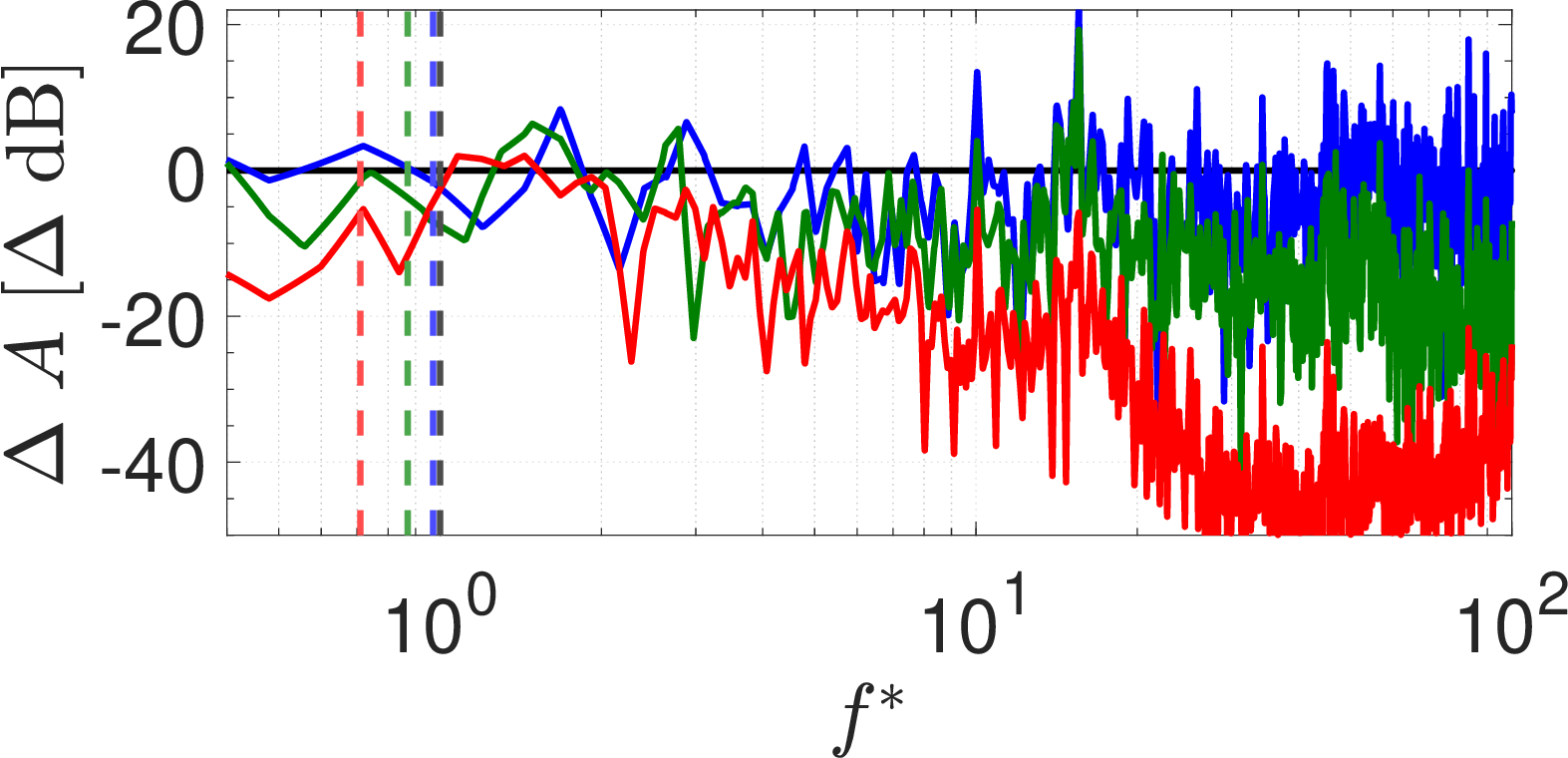}
         \caption{}
         \label{fig:aoa20:fwh:c}
     \end{subfigure}
     \hfill
 \begin{subfigure}[b]{0.45\textwidth}
      \centering
         \includegraphics[width=\textwidth]{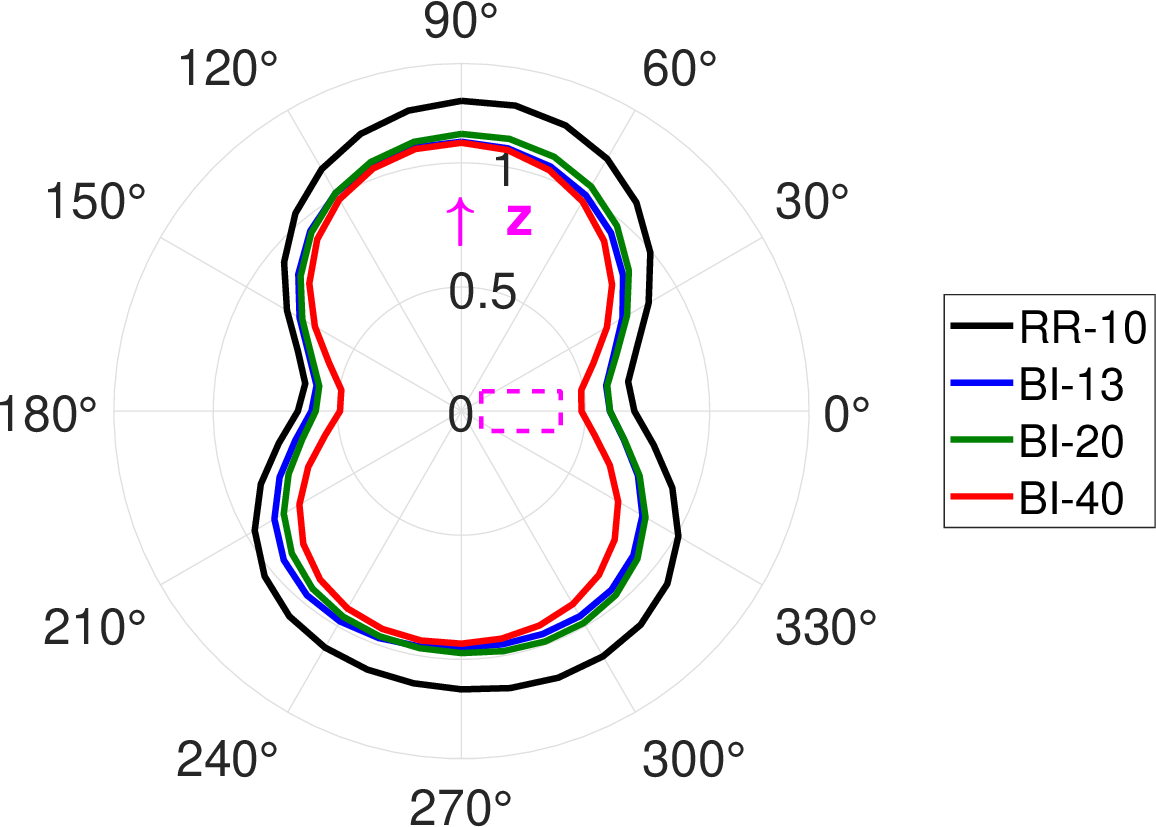}
         \caption{}
         \label{fig:aoa20:fwh:d}
     \end{subfigure}
   \hfill
     \begin{subfigure}[b]{0.48\textwidth}
         \centering
         \includegraphics[width=\textwidth]{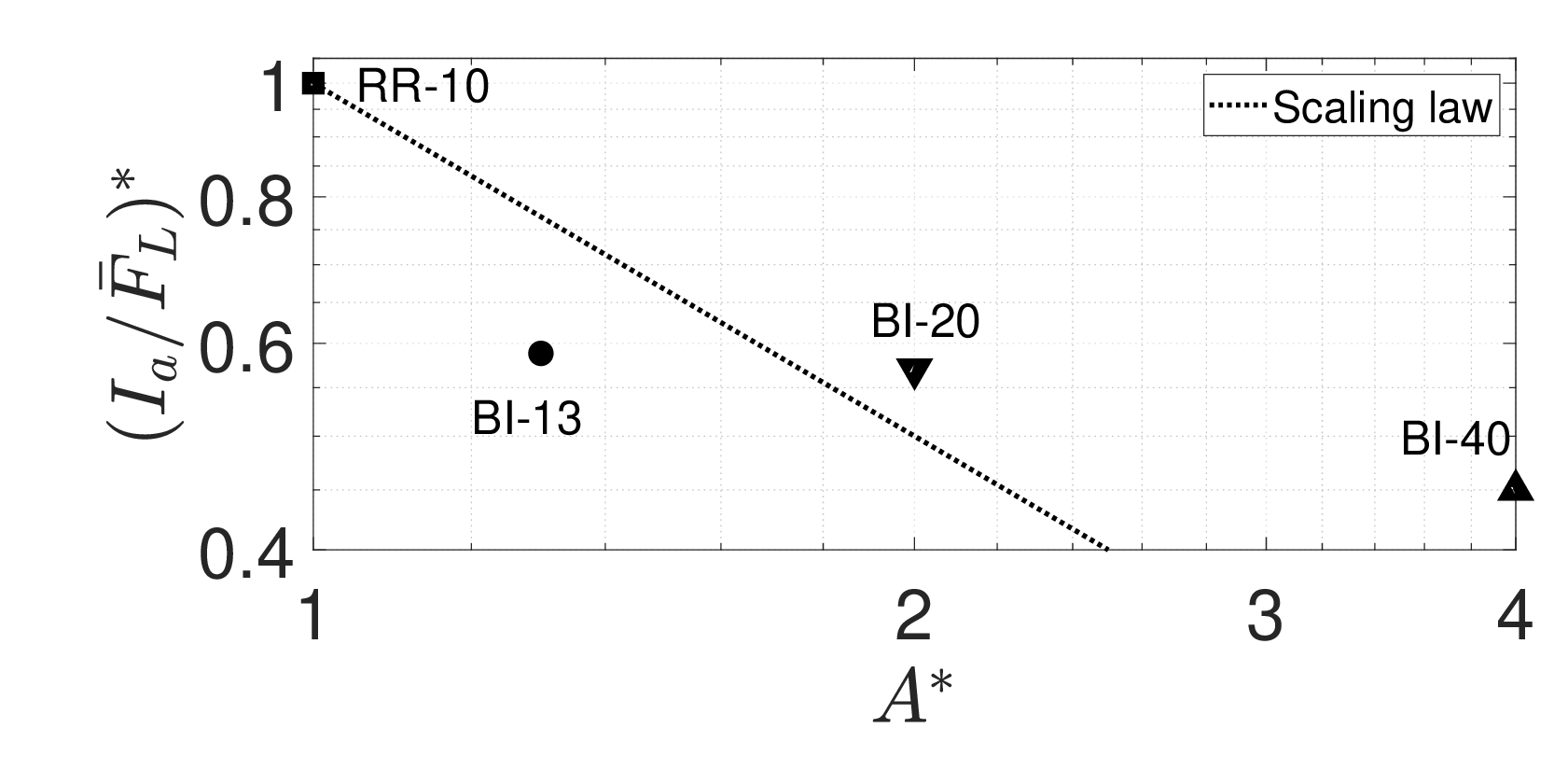}
         \caption{}
         \label{fig:aoa20:fwh:e}
     \end{subfigure}
     \hfill
     \begin{subfigure}[b]{0.48\textwidth}
         \centering
         \includegraphics[width=\textwidth]{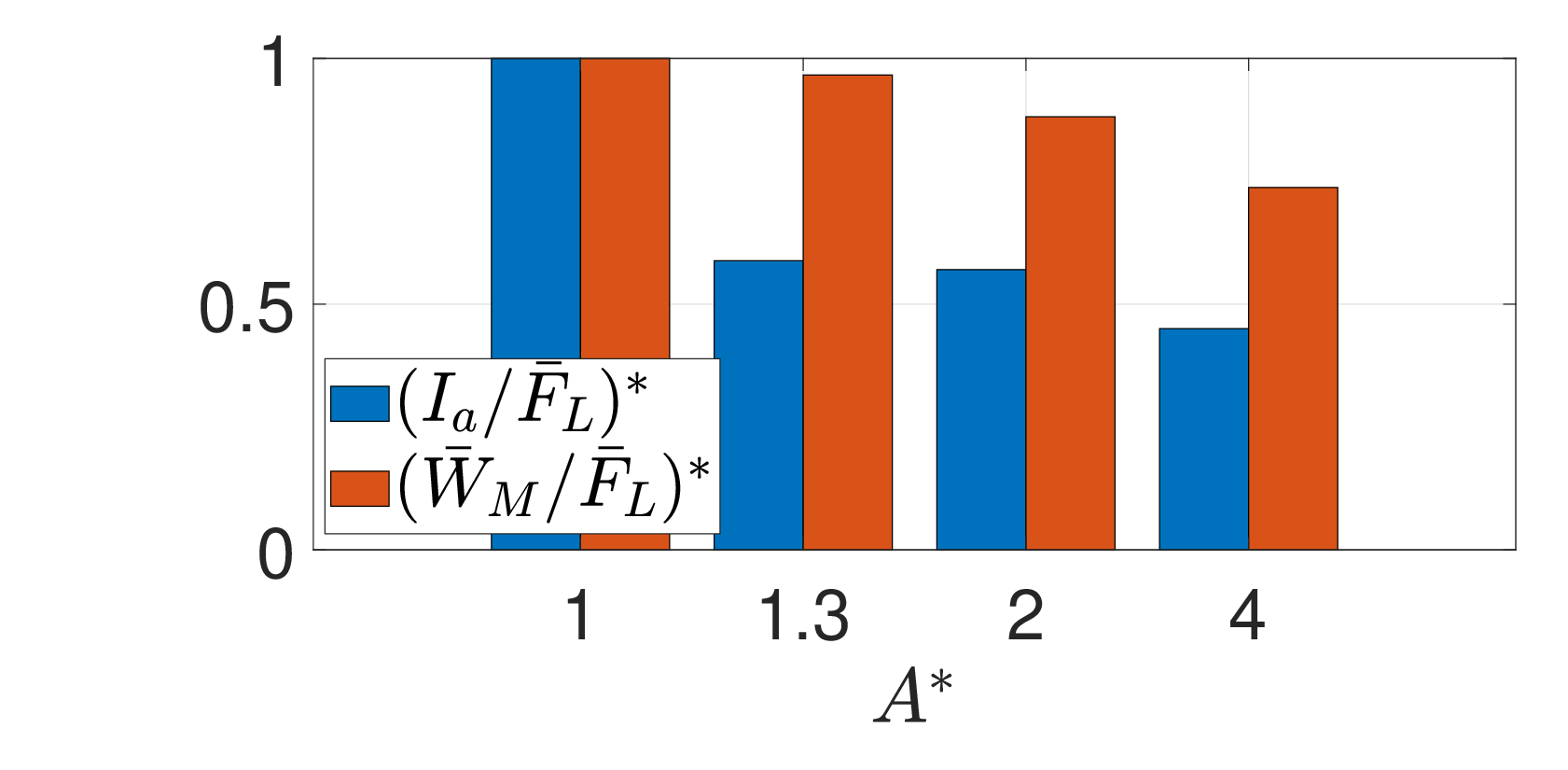}
         \caption{}
         \label{fig:aoa20:fwh:f}
     \end{subfigure}
        \caption{Acoustic results shown for single blade rotors with 20$^\circ$ pitch angle. (a)  The sound spectrum showing the amplitude of sound pressure ($A$) vs frequency ($f^*$) for the RR-10 rotor with frequency normalized using the rotation frequency of the RR-10 rotor. The sound spectrum is shown for all four rotors and calculated at a polar location of (b) (6.5$R_o$, 0$^\circ$) and (c) (6.5$R_o$, 90$^\circ$) with the frequency of the RR-10 rotor used to normalize the frequencies and the amplitude of the RR-10 rotor at each frequency is subtracted from the corresponding amplitude of all the rotors to show the gain or reduction in noise. \oc{The blade passage frequency is shown using dashed vertical lines.} d) The directivity pattern ($p_{rms}^{'*}$) at a distance of 6.5$R_o$ and normalized using the mean directivity corresponding to RR-10 case with dashed pink rectangle showing the blade orientation (not to scale), (e) the scaling of the acoustic intensity with blade area (equation \ref{eqn:intensity_scaling}) and (f) acoustic intensity and mean mechanical power shown for all four rotors. Star superscript indicates that the values are normalized by the corresponding value for rectangular rotor.}
\label{fig:aoa20:fwh}
\end{figure}

The results for the aeroacoustic noise are obtained as before and the noise spectrum for the RR-10 case is shown in figure \ref{fig:aoa20:fwh:a}.  As with the previous cases, the tonal noise peak is very prominent at the $0^\circ$ direction but is missing at the $90^\circ$ location. However, since the drag force is of a lower magnitude than the lift force for this $20^\circ$ pitch case, the corresponding broadband noise at $0^\circ$ is lower compared to the $90^\circ$ location. The directivity pattern is shown in figure \ref{fig:aoa20:fwh:d} and we note that given the larger relative magnitude of the lift force, the directivity exhibits a typical dipole (figure `8') shape.  The scaling of the mean specific acoustic intensity with the rotor area is shown in figure  \ref{fig:aoa20:fwh:e} and we find that the specific acoustic intensity reduces significantly in going from RR-10 to BI-13, but further increase in the area only generated a marginal decrease in this quantity. This is due to two factors; first, the decrease in $C_L$ with increasing area requires that the larger-area blades rotate at a relatively higher speed than for the $45^\circ$ pitch cases. Thus, the reduction in blade tonals, which is expected to scale as $A^{-1}$ does not happen. Second, as noted before, the blade tonals are a smaller component of the noise for these cases and the reduction in broadband noise is not expected to scale with $A^{-1}$.

The results for the aerodynamic power and the acoustic intensity are summarized in figure \ref{fig:aoa20:fwh:f} and also in table \ref{table:20deg_rotor_comparision}and we see that the BI-40 rotor has the best performance out of the four rotors compared here with a 26\% decrease in specific aerodynamic power and 55\% decrease in specific sound intensity. Thus, the strategy of increasing blade area as a means to reduce aeroacoustic noise and aerodynamic power is verified for this case as well.

\subsection{Dual-Blade Rotors}
\label{sec:dual_blade_rotor}
\begin{figure}
    \centering
 \begin{subfigure}[b]{0.32\textwidth}
      \centering
         \includegraphics[width=\textwidth]{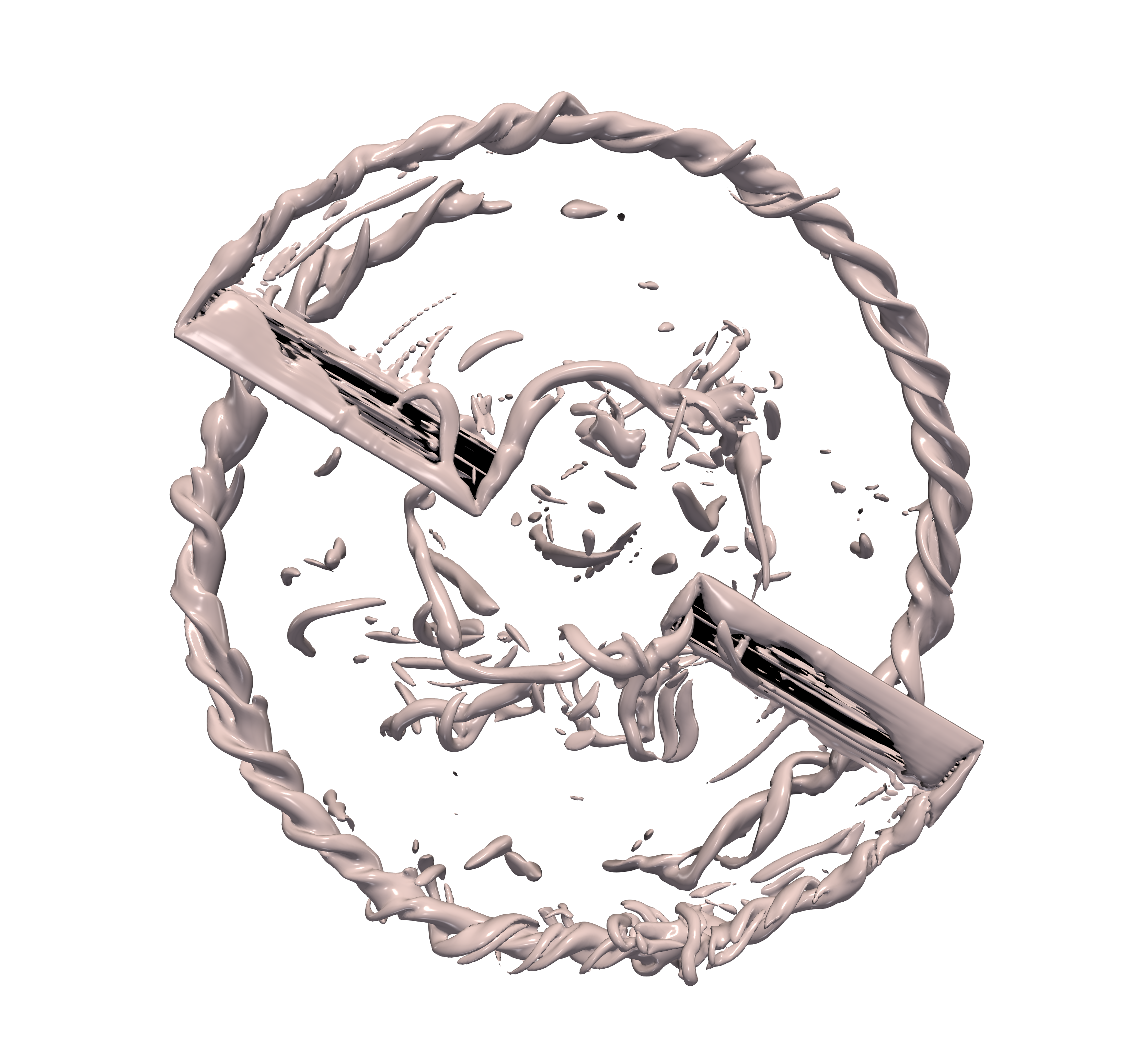}
         \caption{}
         \label{fig:20AoA_double:Qcrit:a}
     \end{subfigure}
  \hfill
     \begin{subfigure}[b]{0.32\textwidth}
         \centering
         \includegraphics[width=\textwidth]{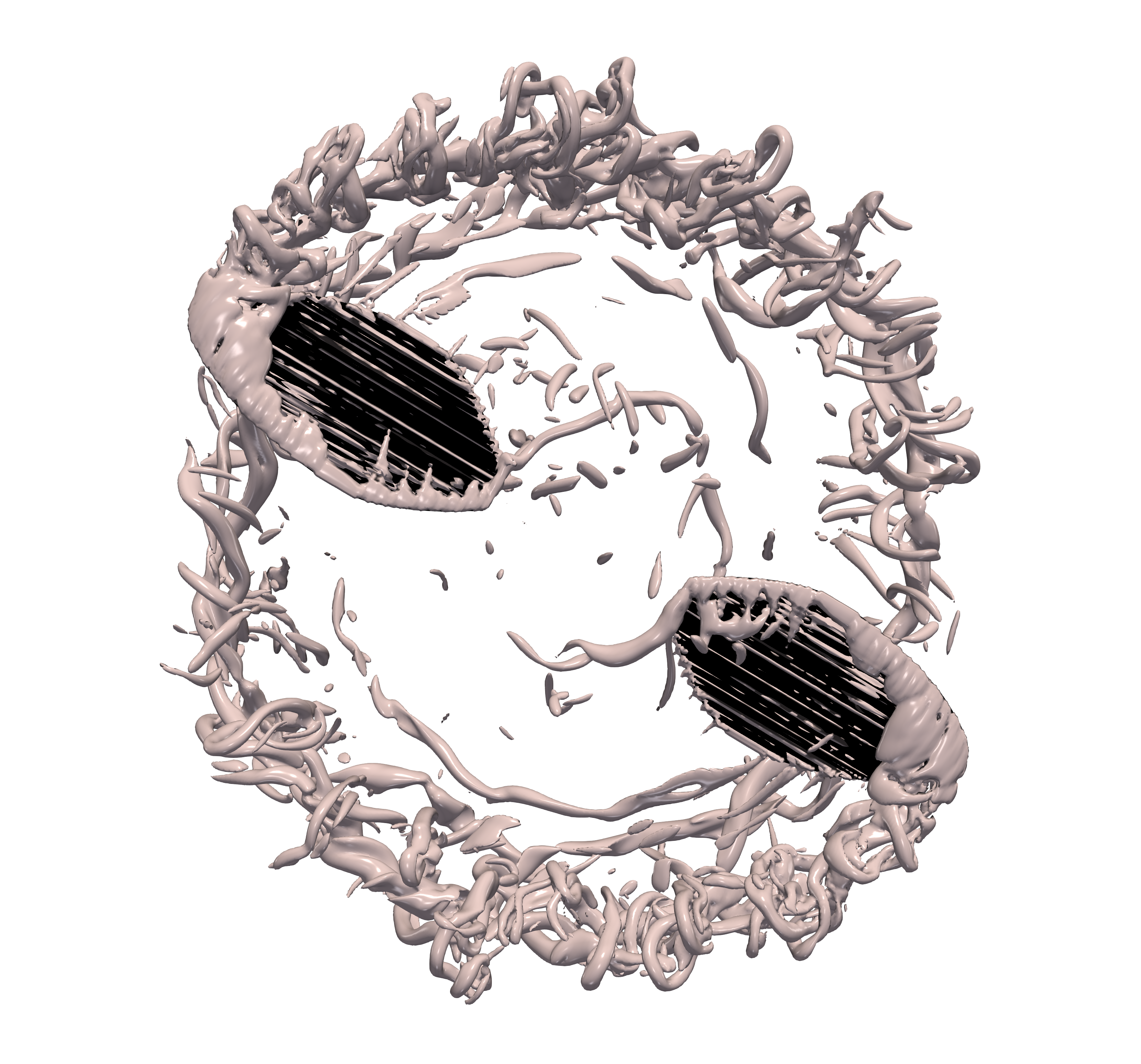}
         \caption{}
         \label{fig:20AoA_double:Qcrit:b}
     \end{subfigure}
     \hfill
     \begin{subfigure}[b]{0.31\textwidth}
         \centering
         \includegraphics[width=\textwidth]{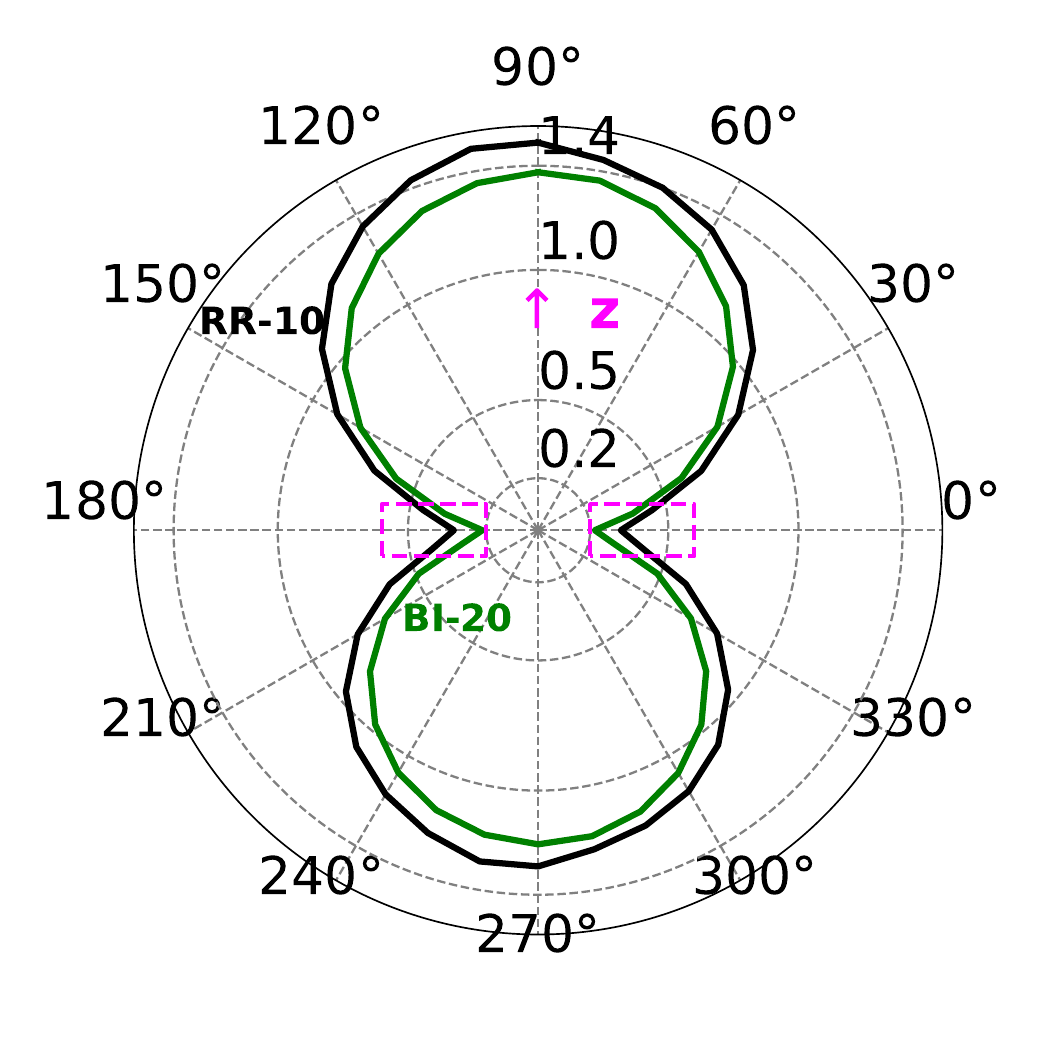}
         \caption{}
         \label{fig:20AoA_double:Qcrit:c}
     \end{subfigure}
     \caption{The vortices shown using the iso-surface of the Q-criterion for the 20$^\circ$ pitch angle dual-blade (a) rectangular rotor RR-10 and (b) the bio-inspired BI-20 rotor. (c) The directivity pattern for these rotors calculated using the RMS value of sound pressure ($p_{rms}^{'*}$) recorded at a distance of 6.5$R_o$ from the center of the rotation and normalized using the mean directivity corresponding to RR-10 case with dashed pink rectangle showing the blade orientation (not to scale).}
\label{fig:20AoA_double:Qcrit}
\end{figure}
In the previous sections, we have investigated a single rotating blade in order to simplify the analysis. While this verified the strategy of increasing blade area as a way to reduce noise and power consumption, single-blade rotors are seldom used and most drone rotors are dual-bladed. In this section, we conduct a limited study of this strategy for a dual-bladed rotor by comparing a dual bladed rectangular rotor RR-10 with the BI-20 rotor. The pitch angle chosen for this study is $20^\circ$, which is more realistic for these rotors. The normalized rotation speed for these RR-10 and BI-20 rotors are 1.0 and 0.88 respectively to produce similar lift force, with the latter found via trial-and-error. \oc{This case was simulated for a total of four rotation cycles with }the snapshot of the vortex structures for these rotors shown in the figures \ref{fig:20AoA_double:Qcrit:a} and \ref{fig:20AoA_double:Qcrit:b}. The RR-10 rotor shows a thin and long spiral shape tip-vortex being formed whereas BI-20 shows a more complex tip-vortex consisting of many vortex loops and it also occupies are larger region of the rotor wake.

\begin{table}
\centering
\begin{tabular}{|l|c|c|}
\hline
\textbf{Quantity} & \textbf{RR-10} & \textbf{BI-20} \\
\hline
$\Omega^*$ & 1.00 & 0.876  \\
\hline
\rb{$\bar{F}_L^*$} & 1.00 & 0.970  \\
\hline
\rb{$\bar{C}_L$} & 0.410 & 0.259  \\
\hline
\rb{$\bar{W}_M^*$} & 1.00 & 0.828  \\
\hline
$I_a^*$ & 1.00 & 0.714  \\
\hline
\rb{$(\bar{W}_M/\bar{F}_L)^*$} & 1.00 & 0.854  \\
\hline
\rb{$(I_a/\bar{F}_L)^*$} & 1.00 & 0.736  \\
\hline
\end{tabular}
\caption{Summary of the mechanical power and aeroacoustic performance results for the dual blade rotors at a pitch angle of 20$^\circ$ and normalized using the baseline case of RR-10 rotor.}
\label{table:20deg_dual_blade_rotor_comparision}
\end{table}
The lift and the lift coefficients are shown in the table \ref{table:20deg_dual_blade_rotor_comparision} \oc{(averaged between last two cycles)} and we note that the difference between the mean lift produced by these rotors differs by about 3\%. Similar to the single-blade rotor at a pitch angle of 20$^\circ$ case, we find a significant (nearly 65\%) reduction in the lift coefficient of the BI-20 rotor compared to the RR-10 rotor.
The directivity pattern shown in the figure \ref{fig:20AoA_double:Qcrit:c} shows the dipole shape and  we find that the BI-20 rotor generated lower noise in all the directions. The mean specific mechanical power and specific sound intensity (calculated at a distance of 6.5$R_o$ from the rotation center) shown in table \ref{table:20deg_dual_blade_rotor_comparision} indicates that the BI-20 rotor requires 15\% less power and generates only 74\% of the noise of the RR-10 rotor per unit lift.

\section{Conclusions}
Inspired by the aeroacoustics of flapping flight in insects, we have examined a simple strategy to reduce the aeroacoustics noise from drone rotors - increased rotor blade area with fixed rotor span, that is accompanied by a concomitant decrease in rotation speed. The study employs direct numerical simulations of the rotor flow coupled with the prediction of aeroacoustic sound using the Ffowcs Williams-Hawkings acoustic model. Four different blades with shapes inspired by insect wings and with different total surface areas are employed. The initial set of simulations are for a single-bladed rotor and we have also examined the effect of blade pitch angle. The study design determines a rotation speed that generated a nearly equivalent mean lift \oc{(and therefore an equivalent disk loading)} for all the different rotors. The key aeroacoustic metric compared between the different blade designs is the aeroacoustic sound intensity per unit lift at a set distance from the center of the rotor and this is found to decrease with increasing blade area. It is also found that an additional benefit of this strategy is a reduction in the aerodynamic power per unit lift required to rotate the rotor. The study culminates with a comparison of a more realistic two-bladed rotor and reductions in aeroacoustic noise intensity and aerodynamic power are found for this case as well. 

The current study has several limitations with the primary ones being the use of non-optimized blade shapes and the inability to directly address non-dimensional rotation rates that are relevant to small-scale drones. These limitations are best addressed via experimental testing and/or the use of simplified models of aeroacoustics that do not require high-fidelity computational models. Notwithstanding these limitations, the current study provides proof that strategies inspired by flying insects can lead to new design ideas for drone rotors that are not only quieter but also more efficient. 

\section{Acknowledgments}
The authors acknowledge support from the Army Research Office
(Cooperative Agreement No. W911NF2120087) for this work. Computational resources for this work were provided by the 
high-performance computer time and resources from the DoD High Performance Computing Modernization Program and Advanced Research Computing at Hopkins (ARCH) core facility (rockfish.jhu.edu).

\end{document}